\documentclass{article}
\usepackage{emulateapj}

\input psfig

\lefthead{Nelson et al.}
 \righthead{Cluster Galaxy Evolution and the LCDCS}

\newcommand\Kp{$K^\prime$\ }
\newcommand\etal{\textit{et al.\ }}
\newcommand\Mstar{M$^{*}_{I}$\ }
\newcommand\mstar{m$^{*}_{I}$\ }
\newcommand\msun{{\rm\,M_\odot}}

\begin{document}
\input{psfig.sty}

%\title{Optical and IR Imaging of 55 Clusters at 0.3 $< z <$ 0.9}
\title{Cluster Galaxy Evolution from a New Sample of Galaxy Clusters
at 0.3 $< z <$ 0.9} \author{Amy E. Nelson and Anthony H. Gonzalez}
\affil{Board of Astronomy and Astrophysics, Univ. of California, Santa
Cruz, CA, 95064, E-Mail: anelson@ucolick.org, anthonyg@ucolick.org}
\author{Dennis Zaritsky} \affil{Steward Observatory, 933 N. Cherry
Ave., Univ. of Arizona, Tucson, AZ, 85721, E-Mail:
dzaritsky@as.arizona.edu} \author{and} \author{Julianne J. Dalcanton}
\affil{University of Washington, Seattle, WA, E-Mail:
jd@toast.astro.washington.edu}

\begin{abstract}

We analyze photometry and spectroscopy of a sample of 63 clusters at
0.3 $\leq z \leq$ 0.9 drawn from the Las Campanas Distant Cluster
Survey to empirically constrain models of cluster galaxy
evolution. Specifically, 1) by combining $I-$band photometry of 44 of
our clusters with that of 19 clusters from the literature
(Aragon-Salamanca \etal; Smail \etal; Stanford \etal) we parameterize
the redshift dependence of $M^*_I$ in the observed frame as $M^*_I =
(-21.74\pm0.12) - (0.88\pm0.24)z - 5 \log h$ (rms deviation = 0.34)
for $0.3 \leq z \leq 0.9$ ($\Omega_{0} = 0.2$, $\Omega_\Lambda = 0$));
2) by combining 30 of our clusters and 14 clusters from the literature
(Aragon-Salamanca \etal; Smail \etal; Stanford \etal) with $V$ and $I$
data we parameterize the redshift dependence of the $V-I$ color of the
E/S0 red sequence in the observed frames as $V-I = (-0.24\pm0.28) +
(7.42\pm1.03)z - (4.61\pm0.91)z^2$ (rms deviation = 0.16) for $0.3
\leq z \leq 0.9$; and 3) by combining 13 of our clusters with 15
clusters from the literature (Aragon-Salamanca \etal; Stanford \etal)
with $I$ and $K^\prime$ data we parameterize the redshift dependence
of the $I-$\Kp color of the E/S0 red sequence in the observed frames
as $I-$\Kp$ = (0.66\pm0.65) + (9.50\pm3.72)z - (14.72\pm7.01)z^2 +
(8.72\pm4.29)z^3$ (rms deviation = 0.18) for $0.3 \leq z \leq
0.9$. Using the peak surface brightness of the cluster detection,
$\Sigma$, as a proxy for cluster mass (Gonzalez \etal), we find no
correlation between $\Sigma$ and \Mstar or the location of the red
envelope in $V-I$. We suggest that these observations can be explained
with a model in which luminous early type galaxies (or more precisely,
the progenitors of current day luminous early type galaxies) form the
bulk of their stellar populations at high redshifts ($\gtrsim 5$) and
in which many of these galaxies, if not all, accrete mass either in
the form of evolved stellar populations or gas that causes only a
short term episode of star formation at lower redshifts ($1.5 < z <
2)$. Our data are too crude to reach conclusions regarding the
evolutionary state of any particular cluster or to investigate whether
the morphological evolution of galaxies matches the simple scenario we
discuss, but the statistical nature of this study suggests that the
observed evolutionary trends are universal in massive clusters.

\end{abstract}


\section{Introduction}

High redshift clusters provide an excellent laboratory with which to
study the evolution of galaxies in dense environments. By investing a
modest amount of telescope time obtaining spectra of a handful of
galaxies, thereby verifying the reality of the cluster and
ascertaining its redshift, one is rewarded with the knowledge of
distances to tens or even hundreds of galaxies. The spectral and
photometric properties of these galaxies can then be used to constrain
evolutionary models, whose differences usually become appreciable at
high redshift, $z > 0.5$ (cf. Bruzual \& Charlot 1993; Kauffmann \etal
1998).  Until recently, such efforts have been hampered by the small
number of known high redshift clusters. Published work in this field
relies upon the examination of the same $\sim$dozen clusters, drawn
primarily from two optical surveys. Both of these surveys (Gunn,
Hoessel, \& Oke 1986, hereafter GHO; Couch \etal 1991, hereafter C91) 
selected objects on
the basis of an overdensity of resolved objects. GHO's efforts yielded
8 clusters with $z \geq 0.55$, while C91 yielded 3.

Because the number of known high redshift clusters is small, they have
been studied in great detail by many authors. In particular, with the
advent of the Keck telescopes and the Hubble Space Telescope (HST),
detailed information regarding the cluster galaxies' star formation
rates and morphologies have complemented the more traditional
ultraviolet, optical, and infrared photometry. As a result, a
comprehensive picture of the formation and evolution of cluster
galaxies has begun to emerge in which many, if not all, galaxies have
experienced some recent star formation. The most active class of
galaxies was first noted by Butcher \& Oemler (1984, B-O). They
studied 33 clusters of galaxies with 0.003$\leq z \leq$0.54 and
discovered that the fraction of blue galaxies increased markedly with
redshift. They interpret the rise in blue fraction as evidence for
increased star formation in the past and this interpretation has
subsequently been confirmed by spectroscopic studies (Dressler \& Gunn
1982, 1983, 1992; Couch \& Sharples 1987; Couch \etal 1994; Abraham
\etal 1996; Fisher \etal 1998; Poggianti \etal 1999). Furthermore,
morphological classifications using high resolution HST imaging show
that the bulk of these star-bursting galaxies are disk dominated
systems, a significant fraction of which show signs of interactions or
mergers (Couch \etal 1994, 1998; Dressler \etal 1994, 1999; Oemler,
Dressler, \& Butcher 1997; Poggianti \etal 1999). 

Cohabitating with these active galaxies, however, is a seemingly
quiescent population. These galaxies are the cluster's brightest and
reddest, which lie along a very narrow locus in the cluster
color-magnitude (CM) diagram referred to as the red
envelope. Detailed morphological and spectroscopic studies of clusters
at high redshift confirm that these galaxies are a subset of the
cluster's elliptical population (e.g. de Propris \etal
1999; van Dokkum \etal 1999). Evolution of the zeropoint, slope, and
scatter about cluster CM relations all suggest that these ellipticals
formed at a very high redshift ($z \gtrsim 5$) and have experienced little or
no recent star formation (see Aragon-Salamanca \etal 1993, hereafter
A93; Rakos \& Schombert 1995, hereafter RS; Ellis \etal 1997; Stanford \etal 1998,
hereafter SED98; Jones \etal 2000), while
work on the evolution of the fundamental plane (Kelson \etal 2000) and
on the evolution of absorption line indicies (Bender, Ziegler, \& Bruzual
1996; Kelson \etal 2001) support this conclusion by placing a lower
redshift limit of $\sim 2.5$ for the formation of the bulk of the
stellar populations. Studies of the optical and infrared
luminosity functions of cluster galaxies suggest a more moderate
formation redshift for cluster
ellipticals, $z \gtrsim 2$ (Smail \etal 1997, hereafter S97; de Propris \etal
1999). Despite this general consensus, the
non-uniqueness of spectral synthesis models, against which the
majority of these observations are compared, 
allows for the possibility that at least some of these galaxies are
not completely
quiescent. In fact, recent work by van Dokkum \etal (1999) suggests
that a large fraction of
these galaxies in the higher redshift ($\sim 1$) clusters are 
not relaxed systems. These authors studied
MS1054-03 at $z = 0.83$ using spectroscopy and HST high resolution
imaging and found that 17\% of the L$>$L$_{*}$ cluster population is
experiencing a major merger, most of which will probably evolve into
luminous (2L$_{*}$) elliptical galaxies. They estimate that $\sim$50\%
of the present-day cluster ellipticals were assembled in mergers since
$z \sim 1$, assuming that the galaxy population of MS1054-03 is
representative. While most of the mergers involve red early-type
galaxies with no detected [OII] $\lambda$3727 emission, some do
exhibit enhanced Balmer absorption indicative of a modest recent
starburst. Consequently, the question of the evolutionary history of
the brightest cluster ellipticals remains open.

As mentioned previously, all of the above work is hampered by the lack
of statistically significant samples. General conclusions based on
small samples rely critically on the assumption that this handful
of clusters is representative of the entire high redshift cluster
population. This assumption is particularly worrisome when (1) it is
viewed in the context
of a hierarchical clustering scenario which holds that structures of
different mass scales form at different times and consequently have
experienced different evolutionary histories, and (2) one realizes that
the cluster properties play a significant role in their original
identification. The need for a large,
well-defined, high redshift cluster sample is evident.

In an attempt to resolve issues raised by small sample size, Postman
\etal (1996) conducted the Palomar Distant Cluster Survey. This survey
consists of optical/near IR imaging of a $\sim$5 deg$^{2}$ area using
the 4-shooter CCD camera on the Palomar 5m telescope. Their cluster
finding technique was more sophisticated than previous optical surveys
by incorporating magnitude and color filters to minimize the
contamination from line-of-sight projections.  Their method relies
upon resolving distant galaxies in two colors and thus requires the
use of a large telescope under photometric conditions with good
seeing, which limits the areal coverage of the survey. Their efforts
yielded a large, well-defined cluster catalog that contains 35
clusters with estimated redshifts $\geq$0.5, plus 25 clusters from an
inhomogeneous sample with estimated redshifts $\geq$0.5 (the PDCS
includes clusters identified previously by GHO). Using these clusters
to study galaxy evolution, their results to date provide more
statistically significant confirmation of the trends observed with
smaller samples. In particular, Lubin (1996) examined the location of
the red envelope in the HST filters $V_{4}-I_{4}$ ($V_{4}$ = F555W
and $I_{4}$ = F785LP) and found that the data are most consistent
with the passive evolution of an elliptical galaxy that formed at high
redshift. She also measured the fraction of blue galaxies, f$_{b}$,
and found an increase in f$_{b}$ with redshift, consistent with a high
redshift extension of the Butcher-Oemler effect.

An alternative, efficient method for finding high redshift clusters
was developed by Dalcanton (1996). This technique capitalizes on the
idea that the total flux from a distant cluster in shallow images
is dominated by the flux from unresolved sources. If one has an intrinsically
uniform image, the flux from the unresolved sources produces a
detectable low surface brightness fluctuation. Because the technique
does not rely upon an overdensity of resolved objects, relatively
shallow (but highly uniform) exposures are sufficient to identify
distant clusters and a much larger area on the sky can be
surveyed. Dalcanton \etal (1997) verified the ability of this
technique to detect LSB objects to the required central surface
brightness using
existing scans from a $\sim17.5$ deg$^{2}$ Palomar 5m survey
(originally taken as part of the Palomar Transit Grism Survey for high
redshift quasars; Schneider \etal 1994). This survey identified more
than 50 northern hemisphere cluster candidates. In 1995, the authors
conducted a 130 deg$^{2}$ drift-scan survey of the southern sky at the
Las Campanas Observatory. The reductions and object identification is
complete and the final cluster catalog contains over 1000 cluster
candidates (Gonzalez \etal 2001). Not only does this catalog result in
a tremendous increase in the number of known clusters at these
redshifts, but because our cluster identification criteria is
independent of those utilized in previous surveys, this catalog
provides an independent, well-defined sample with which to compare the
results of more traditional surveys.

In this paper, we present the first study of galaxy evolution using a subset 
of the Las Campanas Distant Cluster Survey. In addition to 54
southern clusters, we include 9 northern clusters drawn from the
Palomar survey. The combined sample consists of 63
clusters, 17 of which are spectroscopically confirmed and 46 of which
have photometrically estimated redshifts. We obtain deep
follow-up imaging in $V$, $I$, and \Kp, although not every cluster is
imaged in every band.  We use these data, along with that of 19 clusters from
the literature, to classify cluster
candidates, photometrically estimate their redshifts, and study the
luminosity and color evolution of cluster galaxies. In \S2
we briefly describe the survey, including cluster candidate
identification, and our follow-up observations. In \S3 we present our
three photometric redshift estimation techniques. In \S4 we utilize
our clusters with spectroscopic redshifts and those with photometric
redshifts to trace the evolution of cluster galaxies in luminosity and
color. Specifically, we examine the evolution of \Mstar and the location
of the red envelope in $V-I$ and $I-$\Kp. We investigate cluster selection biases and their effect
on our empirically measured galaxy evolution in \S4.4. In \S4.5 we combine our results with those from the recent
literature and suggest a picture of cluster galaxy formation and
evolution. Finally, we summarize our results and conclusions in \S6.

\section{The Data}

Our data originate from a variety of telescopes and instruments.
The candidate galaxy clusters are identified using drift scan images and
techniques described 
briefly below for context, but in full
detail by Gonzalez \etal 2001. We then
describe the follow-up spectroscopy and
photometry of subsamples of candidate galaxy clusters that constitute the 
core dataset for this study.

%\subsection{Selection of Clusters}
\subsection{Cluster Identification}

Our candidate clusters are drawn from two optical drift scan surveys.
The first is a $\sim17.5$ deg$^{2}$ Palomar 5m survey using the
4-shooter camera (Gunn \etal 1987) and the F555W ``wide $V$'' filter
described by Dalcanton \etal (1997).  Dalcanton \etal made use of 
existing scans (originally taken as part of the Palomar Transit Grism
Survey for high redshift quasars; Schneider \etal 1994) to verify the
low surface brightness (LSB) technique for detecting LSB objects to
the required central surface brightness
first proposed in detail by Dalcanton (1996).  This survey identified
more than 50 northern hemisphere cluster candidates.  The second is a
130 deg$^{2}$ survey of the southern sky conducted at the Las Campanas
Observatory 1m telescope over 10 nights by the authors using the Great Circle
Camera (Zaritsky, Shectman, \& Bredthauer 1996) and 
a wide filter, denoted as $W$, that
covers wavelengths from 4500 \AA\ to 7500 \AA. This survey was
specifically designed to
study LSB galaxies and high redshift galaxy clusters. Scans are
staggered in declination by half of the field-of-view (24 arcmin) so
that each section of the survey is observed twice, but with a
different region of the detector. Exposure times are limited by the
sidereal rate, which at the declination of the survey ($-10^\circ$ to
$-12^\circ$) is $\sim$97 sec. The pixel scale is 0.7
arcsec/pixel with a typical seeing of about 1.5 arcsec. The full
region surveyed measures $\sim 85^\circ \times 1.5 ^\circ$ and is
embedded within the area covered by the LCRS (Shectman \etal 1996).  The
result of iterative flatfielding, masking, filtering, and object
detection is a catalog of $\sim$ 1000 cluster candidates (a full
description of the method and catalog are provided by Gonzalez \etal
2001).

The focus of this study is the photometry and spectroscopy of
subsamples of candidate clusters obtained to test cluster
identification algorithms and
study the clusters' galaxy populations. Keck spectroscopy, optical ($V$- and
$I$-band) photometry, and IR ($K^\prime$-band) photometry are
presented, although not all types of data are available for every
candidate.  Table 1 contains information only about those candidates that we 
determine to be bonafide clusters, rather than all observed candidates.
The name of the cluster is listed in the first column.  RA and DEC (2000.0) are
in the next column, where the
center of the cluster is taken to coincide with the centroid of the low
surface brightness fluctuation. 
The last six columns contain the
exposure time and seeing for each of the three photometric
bands, $V$, $I$, and \Kp.  In summary, 97$\%$ of our clusters are imaged in $I$, $54\%$ in $V$,
$33\%$ in \Kp, and $29\%$ have spectroscopy.
Some clusters were imaged twice in various photometric bands. Because
objects were observed on different runs (sometimes with a
different camera and/or telescope) and calibrated using
different sets of standard stars, they often have two completely independent
sets of photometry. We use both sets independently in the
calibration of our photometric redshift estimators and subsequent
analysis.

These data provide a set of
confirmed clusters with which to develop a set of quantitative
statistical descriptors to use for cluster selection and a statistical
sample of clusters with which to study galaxy evolution in clusters.
Because we were testing our cluster selection algorithms, the cluster subsample is heterogeneous. It includes both some of our
best and some of our worst candidates.
Because our selection criteria evolved during this study, 
some of the confirmed candidates presented here will not be in our final
cluster catalog.
No biases in galaxy properties, except those
that possibly correlate with cluster properties such as concentration
and richness (parameters that factor into the cluster selection),
are expected. This issue will be discussed in more detail in \S4.

\subsection{Spectroscopy}

Two critical cluster parameters for any subsequent study of our catalog
are the cluster redshifts and the masses (or alternatively to mass, 
richness, X-ray luminosity, or velocity dispersion, albeit with
different caveats). 
With a catalog of 
$\sim$ 1000 cluster candidates, it is not feasible to obtain spectroscopy
or deep follow-up photometry for every candidate. Consequently, it is imperative to develop
efficient observational estimators of these parameters. 
We present and evaluate several redshift estimators.

In December 1995, we began an observing program that combines long-slit
spectroscopy obtained 
using the Keck telescopes and optical and infrared photometry
obtained using the Las Campanas 2.5m and 1m telescopes, the Palomar
1.5m, and the Lick 3m. 
Our basic goal is to identify and calibrate
photometric redshift estimators. These data have two additional
functions: 1) they enable us to confirm candidate clusters and tune our selection criteria, and
2) they enable us to study the clusters themselves
and their galaxy populations.

Spectra were obtained using the Low-Resolution Imaging
Spectrograph (Oke \etal 1995) 
on the Keck I telescope on 1995 December 20-21, 1997 March 14-15,
1998 April 4-5, and 1999 March 20-21. Due to weather and technical
problems $\sim$ 40\% of the time was usable.
The deep imaging necessary to 
construct multi-object masks was generally not available so 
we used a single long slit with a 600
line mm$^{-1}$ grating.  The slit was aligned, using the guider image, 
to include as many individual galaxies as possible (generally 2 to 3
with m$_{R} \lesssim 22$) and to lie across the position of the LSB feature
detected in the scans.  
As an interesting aside, the Keck guider images are deeper than
the drift scan images from which the candidates are selected.
Typical spectroscopic exposure times were 30
minutes to one hour. Preliminary results from this work
are discussed by Zaritsky \etal (1997).

The reduction of these spectra is standard.
Images are rectified and calibrated using
calibration-lamp exposures and night sky lines (see Kelson \etal
1997 for details).  Velocities are measured using a cross-correlation
technique or by measuring centroids of emission lines.  The rms redshift
difference between redshifts measured from different emission lines
in the same spectra is 0.0033 (typically less than 0.0005).  In cases
where only one emission line is observed, we attribute it to H$\beta$.
Little ambiguity exists between H$\beta$ and [OII] because the [OII]
emission line is a resolved doublet (3726 and 3729 \AA).  

\begin{figure}
\plotone{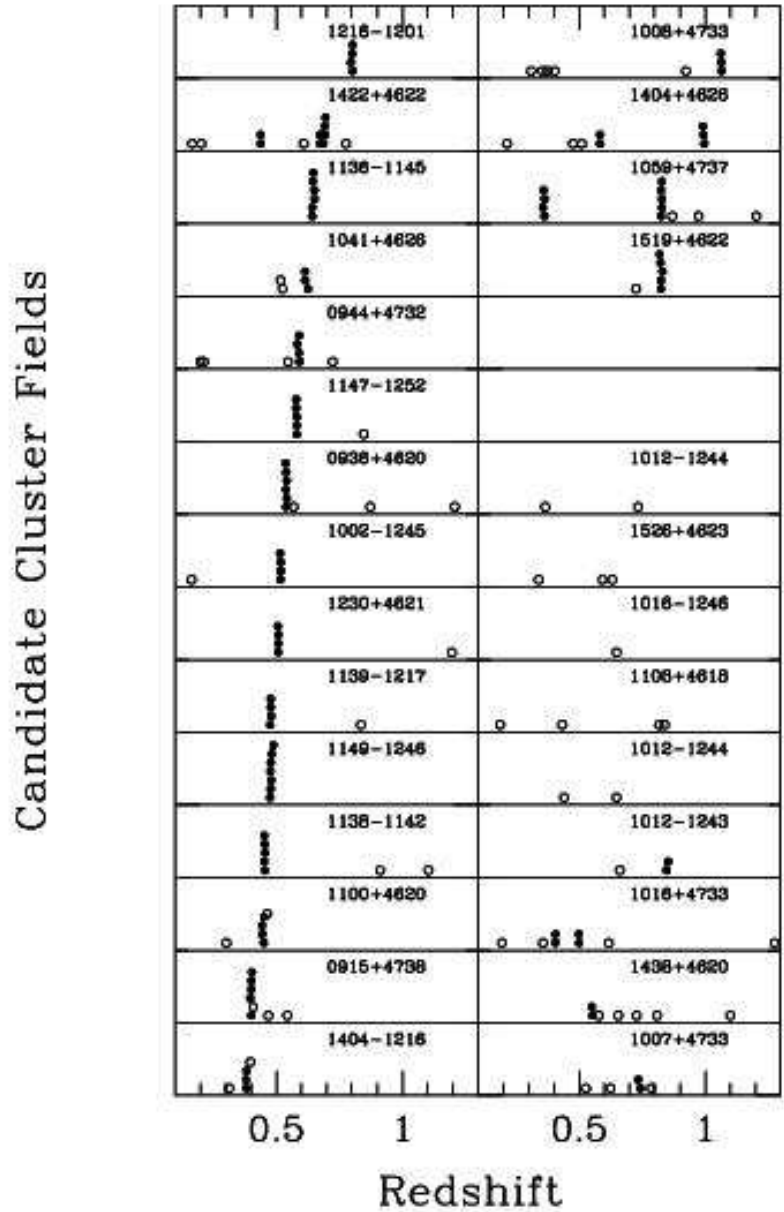}
\caption{The redshift distribution of galaxies in all cluster fields
observed with LRIS on Keck. Each panel represents the results for one
candidate cluster field. Each circle represents one galaxy redshift.
Filled circles represent galaxies within 1000 km sec$^{-1}$ of
another galaxy. Circles are stacked vertically within each panel to 
avoid overlap in the Figure. The two empty panels do not represent
candidate cluster fields.}
\label{fig:z_dist} 
\end{figure}

The distribution of galaxy redshifts is shown in Figure~\ref{fig:z_dist} 
for all 28 fields observed.  
We consider a candidate object to be a real cluster if at least three
galaxies cluster within 1000 km s$^{-1}$ of their mean redshift. A
valid concern is that ``successful" candidates in this scheme may
arise randomly in lines-of-sight through large-scale structures. If
random large scale structures appeared in 19 of 28 fields (60\%), then
we would expect to find two coherent structures in 13 of our
fields. Only one such field exists, and so it is evident that random
redshift associations are not the {\it dominant} cause of the redshift
associations. However, to test whether random redshift associations
are {\it significant}, we use published Keck LRIS spectroscopy 
(Guzman \etal 1997; Lowenthal \etal, 1997; Phillips \etal 1997;
Vogt \etal 1997; Cohen \etal 2000) out to $z = 1.1$ to
estimate the likelihood of chance redshift associations. The use of
data from the same instrument as used in this study ensures that the
data are of comparable quality and have similar systematic problems
(sky lines, fringing) over the same redshift range. The joint catalog
contains coordinate positions and redshifts for 686 galaxies at $0.3 < z < 1.1$ (a redshift range chosen to match
 that of our sample). We then randomly select fields with
the same distribution of galaxies per field as seen in Figure~\ref{fig:z_dist}.
The number of pairs of galaxies within 1000 km sec$^{-1}$ predicted for
our sample from 10,000 simulations is 9.9 (we observe 7, which is
within the 1$\sigma$ Poisson fluctuation). The number of triplets
predicted is 1.3 (consistent with the one system in which two
structures are seen, although this is not necessarily the spurious system).
Only four of our successful 19 cluster candidate have the minimum
three members. The number of quadruples predicted in this sample is
0.2 (we observe 15).  We conclude that systems with three or more
galaxies within 1000 km sec$^{-1}$ are overwhelmingly not the result
of random lines of sight through large-scale structures. This
conclusion will later be confirmed by the independent correspondence
of the magnitude of the brightest galaxy in each cluster candidate,
the mean spectroscopic redshift of the galaxies in the overdensity,
and the concentration of galaxies at the position of the original
LSB detection. We
conclude that our success rate for identifying true clusters is
$\gtrsim$ 60\%, and that {\it perhaps} it is substantially
greater. This success rate will be confirmed independently by other
means below.

\subsection{Optical Photometry}

$V$- and $I$-band images, which straddle the Ca II H \& K
break for galaxies at redshifts of $0.3 \lesssim z \lesssim 0.9$, provide
a measure of the stellar populations in these galaxies.
Because 
the majority of our photometry was completed before the cluster finding 
algorithm was fine-tuned (indeed it was necessary to help train the algorithm),
we imaged some candidates only in $I$, to provide more images of 
potential clusters at the expense of color information.  
Optical photometry in the Johnson $V$-band and Cousins $I$-band for the
northern hemisphere clusters was obtained at the Palomar 1.5 m
telescope during 1996 May 9-14.  Conditions during this run were
rather poor due to intermittent cirrus with seeing ranging from
2$^{\prime\prime}$ to 3$^{\prime\prime}$.  Typical exposure times were
between 0.5 and 1.5 hr.
The southern hemisphere clusters were imaged
in $V$ and $I$ at the Las Campanas Observatory using the 1 m (1997
February 26-28 \& March 1-7, 1998 March 27-31 \& April 1) and 2.5 m
(1996 March 24-26, 1997 April 11-15, and 1998 March 28-29) telescopes.
Conditions during these runs varied, but were
predominantly photometric with seeing typically
$\sim$1$^{\prime\prime}$ (often sub-arcsecond) on the 2.5 m and
$\sim$1.3$^{\prime\prime}$ on the 1 m.  Exposure times ranged between 0.75
and 1.5 hr on the 1 m and 0.3 and 1 hr on the 2.5 m.  

Data reduction is once again fairly standard. Individual
frames are bias subtracted, flatfielded with either twilight or dome
flats, and combined with appropriate high and low sigma clipping to
remove cosmic rays and dead pixels.  Calibration is done using
Landolt's (1983, 1992) standard fields.  All fields observed at
Las Campanas were observed
at least once during photometric conditions.  Although the nights
during the Palomar run were generally not photometric and bootstrapping
photometric exposures are not available, the
solutions for the standards taken during the run show only modest
scatter (0.06 mag) that suggests the conditions were reasonable.  
Nevertheless, those data should be viewed with caution.

Galaxy magnitudes are measured using SExtractor's (Bertin
\& Arnouts 1996) ``best'' total magnitudes.  For uncrowded objects,
this is the automatic aperture magnitude, which 
misses less than 6\% of the light from a galaxy. The corrected
isophotal magnitude, which misses less than 15\% of the galaxy's light
but is more robust than the automatic aperture magnitude,
is used for crowded objects. Although the internal uncertainties in the
SExtractor photometry almost always dominate, we include the
uncertainty from the photometric solutions in determining the uncertainty in
the final magnitudes. Galaxy magnitudes are corrected for extinction
using the dust IR emission maps of Schlegel \etal (1998), but are not
K-corrected. 

\subsection{IR Photometry}

We image a subset of clusters in the infrared \Kp band (2.16$\mu$m) to
probe the older, underlying stellar populations.  A related advantage
is that K-corrections are smaller than for the optical passbands,
simplifying translations to the rest frame magnitudes.  The principal
disadvantage is decreased efficiency due to the bright sky
background at $K^\prime$.

Once again, we used several telescopes to collect the data.  The
northern hemisphere clusters and a few southern hemisphere clusters
were observed at the Palomar 1.5 m telescope (1996 January 7-12), and
the Lick 3 m telescope (1996 April 26-29 and 1998 March 6-9).  A
majority of the nights were photometric, although the seeing was
rather poor (1$^{\prime\prime}$ to 2$^{\prime\prime}$).  Exposure
times, on object, were typically 1 to 2 hr at Palomar and 0.6 to 1 hr
at Lick. We observed the bulk of the southern clusters at the Las
Campanas 2.5 m telescope during 1996 March 31 and 1996 April 1-2.
Every night of this run was photometric with excellent seeing
(0.6$^{\prime\prime}$ to 1$^{\prime\prime}$). Typical exposure times
were 0.5 to 0.75 hr on object. The telescope was always dithered to nearby
blank fields at frequent intervals to monitor the rapidly changing
sky.

Data reduction follows the standard IR procedure. We subtract the dark
current from individual images, flatfield using either twilight or
dome flats, and subtract the sky.  Images are aligned and combined
with an appropriate sigma clipping to remove cosmic rays, and hot or dead
pixels that were not masked.  Every cluster with IR photometry was
observed at least once during photometric conditions.  Calibration was
performed using the HST faint near-IR standards (Persson \etal
1998).  The photometric solutions for the standards taken during the
1996 Lick run show large scatter (0.1 mag) either because the nights
were not photometric or because only a few observations of standard
stars were made. Consequently, the photometric solutions are not
sufficiently well determined that one or two outliers can be
confidently removed. \Kp magnitudes were measured using SExtractor's
(Bertin \& Arnouts 1996) ``best'' total magnitudes (see above for a
description).  As with the optical photometry, the uncertainty in the
final magnitudes is determined by folding in the error in the
photometric solutions with SExtractor's internal photometric
uncertainty. Again, the galaxy magnitudes are corrected for extinction
(Schlegel \etal 1998), but are not K-corrected.

\section{Photometric Redshifts}

The first step in any potential study of the galaxy cluster candidates
is to obtain redshift measurements.  Obtaining spectroscopic redshifts
for a sample as large as ours is not feasible, so we must construct a
reliable, but efficient photometric redshift technique.  We develop
and test three techniques. First, we use the observed magnitude of the
brightest cluster galaxy (BCG) measured on our drift scan images to
obtain $z_{BCG}$. Second, we use the $I-$band luminosity function
measured from the deeper $I$-band follow-up observations to obtain
$z_{lum}$. Third, we use the location of the ``red envelope'' in the
cluster color-magnitude (CM) diagram measured using the $V$- and
$I$-band follow-up observations to obtain $z_{col}$.  We empirically
calibrate all three methods using a subset of our clusters with
spectroscopic redshifts. In addition, we include 19 clusters taken
from the literature (A93; S97; SED98) in
the calibration of $z_{lum}$ and $z_{col}$. 
An empirical calibration is advantageous because the estimated redshifts are independent
of assumptions of cosmology or galaxy evolution. Recall that several
of our clusters were observed twice in various bands on different
runs. Because they have independent calibration and photometry, we use
both data sets in our calibration.

The methods vary in robustness and efficiency.  The most
observationally efficient is $z_{BCG}$ because no follow-up
observations are necessary (aside from those required to calibrate the
method). We can measure the magnitude of the BCG directly from the
drift scan images out to $z \sim 0.8$.  Consequently, this is the only
redshift indicator that we have for all of our cluster candidates with 
$z \lesssim 0.8$.
However, it would appear to be the least robust option because it
depends on the proper identification of one galaxy and on the
homogeneity of BCG luminosities at a given redshift.  Fitting the
galaxy luminosity function of each cluster and using \Mstar as a
standard candle circumvents issues related to small number statistics
and the proper identification of the BCG. Unfortunately, other
difficulties arise.  The cluster galaxy catalog must be corrected for
contamination from interlopers by statistically subtracting the
background galaxies, which introduces a significant source of
uncertainty. Furthermore, the method is intrinsically less efficient because it
requires follow-up imaging of each cluster.  The third method, which
utilizes the location of the red envelope, does not require background
subtraction because the red ellipticals in the cluster are typically
the reddest objects in the field, but requires follow-up observations
in two separate filters.  Although no method appears perfect, they
provide multiple cross checks against each other, and against the
spectroscopic redshifts. As we show next, they all work remarkably
well.

\subsection{Estimating Redshifts Using BCGs}

BCGs are excellent standard candles at low
redshift, $z \lesssim $0.05, with 
absolute magnitude dispersion of 0.30 mag in $V$ and 0.24 mag in $R$,
when corrections
based on environment are applied (e.g. Sandage \& Hardy 1973; Hoessel,
Gunn \& Thuan 1980; Hoessel 1980; Schneider, Gunn, \& Hoessel 1983a,
1983b; Lauer \& Postman 1992; Postman \& Lauer 1995).  More
recently, their near constancy in luminosity has been shown to extend to
$z \sim $1 in the $K$-band (A93; Aragon-Salamanca \etal 1998;
Collins \& Mann 1998).  Because of this stability, their large intrinsic
luminosity, and the fact that we can measure their magnitudes directly
from the drift scan images, they are the most accessible photometric
redshift indicator available to us. 

We define the BCG to be the brightest galaxy within 15$^{\prime\prime}$
of the candidate cluster center.
A fundamental concern is that BCGs may not necessarily lie coincident
with the galaxy luminosity centroid measured by our LSB technique.
Although this is a valid concern, 
the overall success of the method argues that while any particular
cluster may be affected, it is not a general problem.
(This issue is discussed in more detail by Nelson \etal 2001.)
To obtain the BCG magnitude, we measure the light within
a 5$^{\prime\prime}$ aperture using 
SExtractor. A second valid concern is that a fixed angular aperture, which
will correspond to different physical scales at different redshifts,
is used to determine the magnitude. However, 
because we calibrate the magnitude-redshift relation empirically, 
we are only assuming that at \textit{each} redshift the population is uniform (any
bias introduced by using different metric apertures at different redshifts
will calibrate itself out of the relation).

\begin{figure*}
\plotone{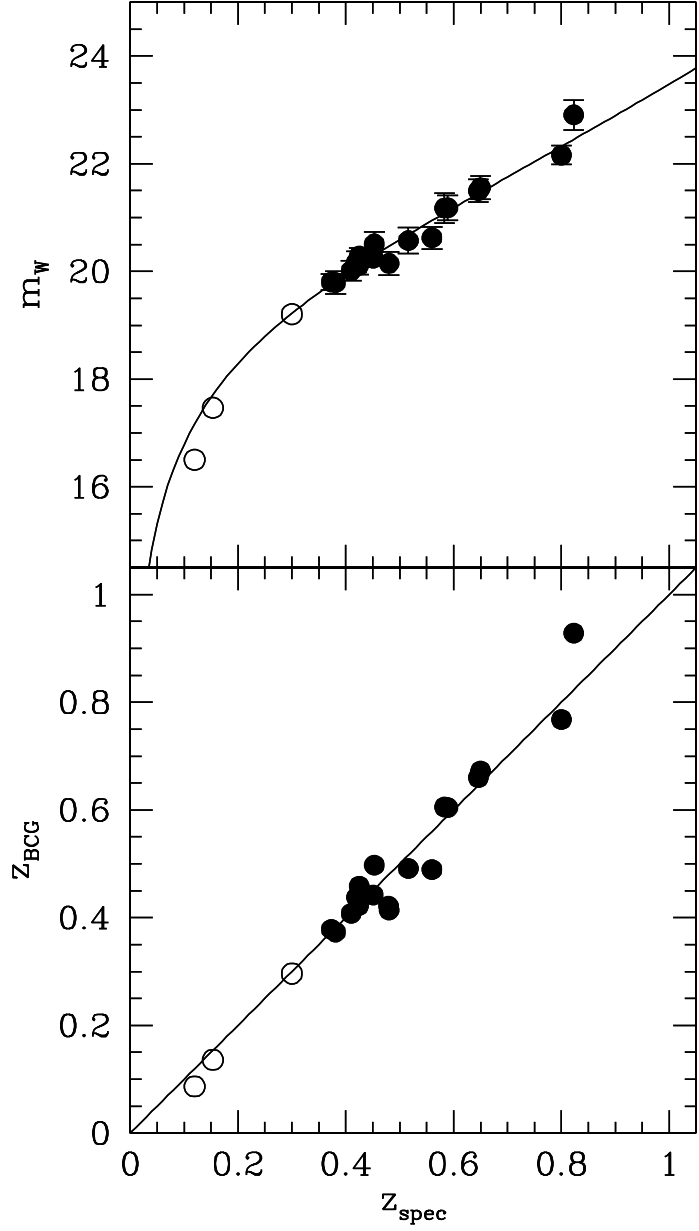}
\figurenum{2}
\caption{\textit{upper panel:}  The derived BCG
magnitude observed in the $W$ filter, m$_{W}$, vs. $z_{spec}$
for clusters from our survey (\textit{filled circles}) and three known low redshift
clusters that fall in our cluster survey area and have redshifts
published in the literature (\textit{open circles}).
Error bars represent the photometric errors and errors due to the surface
brightness correction term, summed in quadrature.  The solid line is our parameterization of this
relation (\textit{see text for details}).  \textit{lower
panel:} Comparison of $z_{spec}$ vs. $z_{BCG}$.  The line is the
expected 1:1 correlation of $z_{spec}$ and $z_{BCG}$. The data have an rms
dispersion of 0.05.\label{fig:bcgs}}
\end{figure*}

We use a subset of 17 clusters\footnote{We cannot reliably 
measure the cluster surface brightness of about 1/3 of the cluster candidates
identified in our earliest analysis because those objects are near
bright stars. Subsequent catalogs (including our final catalog presented
by Gonzalez \etal 2001) are drawn from analyses using larger masked
regions around bright stars and galaxies.
Because we utilize a surface brightness correction in our
m$_{BCG}$ vs. $z$ relationship,
there are two clusters with spectroscopic redshifts
that are not included in the calibration.} with spectroscopic redshifts from our survey to calibrate
the magnitude-redshift relation, shown in the upper panel of
Figure~\ref{fig:bcgs}. We model this relationship with a function that includes 
both luminosity evolution and a second-order correction term based on the 
surface brightness ($\Sigma$) of the cluster in the detection data (see 
Gonzalez \etal 2001) to constrain the function to have the generally understood behavior
of distant BCGs. 
Specifically, we parameterize the magnitude-redshift 
relation as
\begin{equation}
m_{obs}=M+5\log_{10} (D_L/10 {\rm pc}) -A \log10((\frac{\Sigma}{10^{-2}})(\frac{1+z}{1.5}))^\alpha + B z^\gamma.
\end{equation}
The third term is the surface 
brightness correction which decreases the scatter by a factor of $\sim$2. 
The last term parameterizes evolution and is based on results from Bruzual
\& Charlot models (Bruzual \& Charlot 1993). Because the magnitude changes
nearly linearly with redshift, the nature of the parameterization is
not critical (a linear function of $z$ is on average only different by
0.01 mag from our adopted relation). Here, and throughout this
work, we work in the observed frame, not the rest-frame, and as such,
any ``evolution'' will include both real evolution in the stellar
populations plus K-corrections. The free parameters 
in this fit are $A$, $B$, $\alpha$, and $\gamma$ (see Gonzalez \etal 
2001 for a full discussion).  We 
fix the cosmological parameters $\Omega_{M} = $0.3 and 
$\Omega_{\Lambda} = 0.0$ for computing $D_{L}$.  This choice of cosmology 
has no impact on the results, as we are only interested in using this 
empirically calibrated relation to estimate redshifts within the redshift 
range spanned by the calibration data.

This parameterization of the relationship works well.  
A comparison between $z_{BCG}$
vs. $z_{spec}$ is presented in the lower panel of Figure~\ref{fig:bcgs}.
The rms about the 1:1 line is 0.05.  Because the errors in $z_{spec}$
are negligible in comparison to those in $z_{BCG}$, we adopt
$\sigma_{z_{BCG}} = $0.05.  In addition to our high redshift
cluster sample, there are three previously
known clusters that fall within our survey area and have published redshifts. 
We measure their BCG
magnitudes from the drift scans (open circles in 
Figure~\ref{fig:bcgs}) and
find that they are in remarkably good agreement with our
parameterization of the m$_{BCG}$ vs. $z$ relation, even though they lie
outside the redshift calibration range. For all of our cluster
candidates with detected BCGs, we invert this
relation to estimate a cluster redshift. Uncertainties due to
misidentification of the BCG or on the the measurement of $\Sigma$ are
discussed by Gonzalez \etal (2001) who conclude that $\sigma_{z_{BCG}}$ is
0.08. The quoted uncertainties on the photometric redshift indicators may be
underestimates of the uncertainties for the catalog as a whole, because the spectroscopic
clusters are generally richer cluster candidates, which may have more
homogeneous BCGs.

\subsection{Estimating Redshifts Using the Cluster Galaxy Luminosity Function}

The galaxy luminosity function can be characterized using a
Schechter (1976) function, with two free parameters \mstar and $\alpha$.
Our photometric redshift estimation technique relies on
$\chi^{2}$ minimization of this functional form to our $I$-band
data to measure \mstar.
The $I$-band data are less sensitive to stellar population differences
among galaxies than
bluer optical bands, and much more observationally efficient
than IR bands. Because our
data at best reach only a few magnitudes below \mstar, we fix
the slope of the faint end of the luminosity function at $\alpha = -1.25$ 
(Lugger 1986) and only use galaxies with m$_{I} \lesssim 23$ in the
fit. 

\begin{figure*}
\figurenum{3}
\plotone{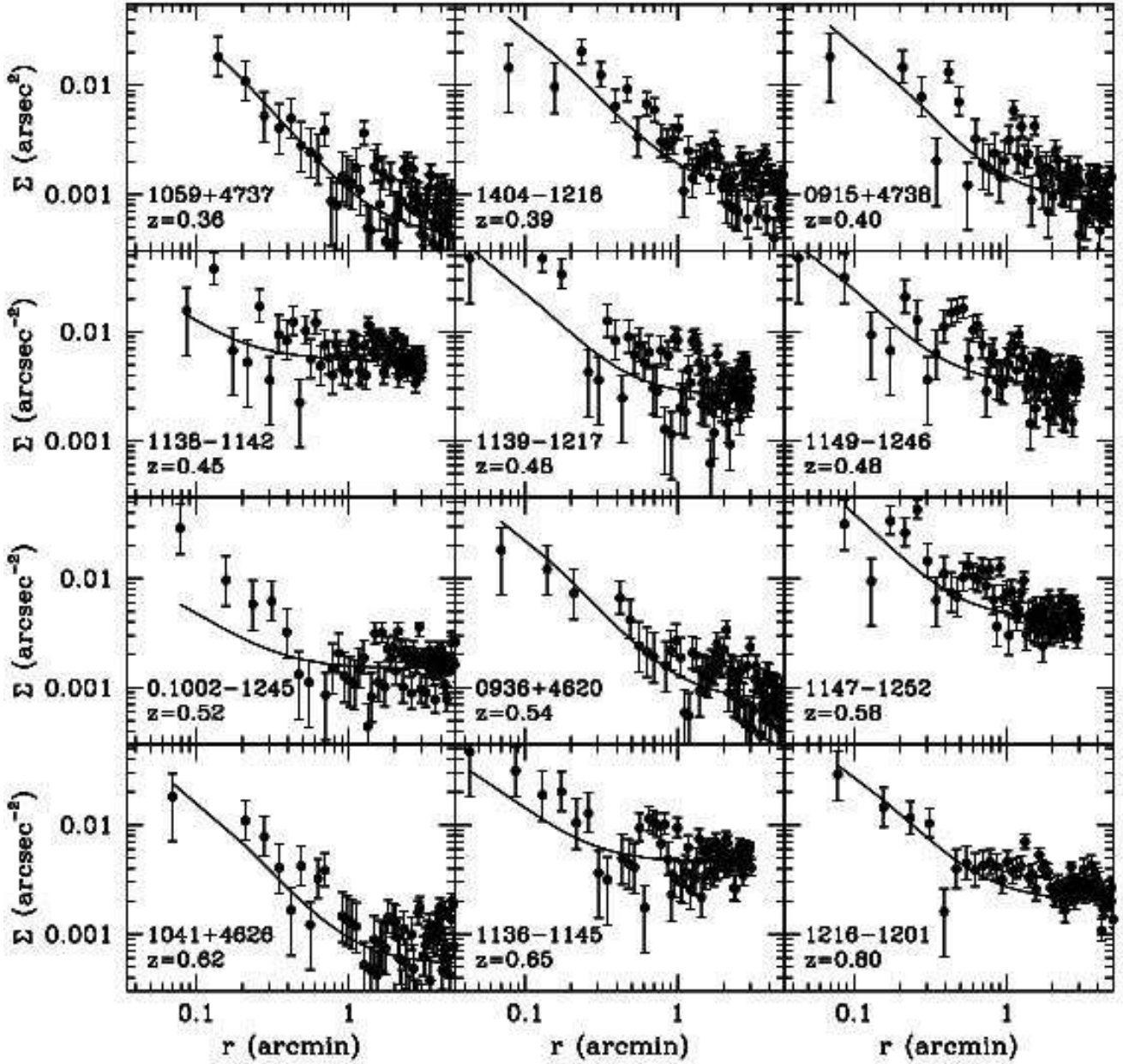}
\protect\figcaption{The surface number density profiles for our
spectroscopically confirmed clusters with $I$-band luminosity
functions. All galaxies within 700 pixels of the cluster center are
considered. We fit the profiles with a radial profile of the form
$C_{1} + C_{2}/(1 + r)^{2}$ (\textit{solid lines}).}
\label{fig:radplot}
\end{figure*}

To generate the cluster luminosity function, we define a
cluster center and radius, and remove contamination due to projected
stars and interloping galaxies.  
The
initial cluster center is taken to coincide with the centroid of the LSB
fluctuation in the drift scan detection.  
The centroid of the positions of the 30 galaxies closest to
the initial cluster center is used as the final center.  
The overall shift from initial
to final center is typically less than 20
pixels ($\lesssim10^{\prime\prime}$).  We fit the surface number
density of galaxies within 700 pixels of
the cluster center with a radial profile of the form
$C_{1} + C_{2}/(1 + r)^{2}$.  The surface number densities for our 12
spectroscopically confirmed clusters with $I$-band luminosity
functions are shown in Figure~\ref{fig:radplot}. The initial cluster radius is
defined to be the radius at which the surface density of galaxies is
10\% greater than the background surface density, where the background is
defined at $r > r_{cl} + 100$ pixels, and $r_{cl}$ is the cluster radius. Because our data are primarily
sensitive to the bright cluster galaxies that lie in the central
regions of the clusters, the radial fits yield a rather crude
estimation of the cluster radius. Therefore, the initial
cluster/background radius is incrementally increased by 10\% until 
an increase in radius adds more
background galaxies than cluster galaxies.  We typically become dominated
by background galaxies at $r_{cl} \sim 150-300 h^{-1}$ kpc and
consequently our luminosity functions are derived from central cluster
galaxies only. Stars are removed from
the object catalog using SExtractor's star/galaxy separation index, CLASS$_{-}$STAR.  To account for the effect of 
differing seeing conditions on the classification
parameter, we examine the behavior of peak
surface brightness of objects vs. CLASS$_{-}$STAR, magnitude vs. CLASS$_{-}$STAR,
and magnitude vs. peak surface brightness.  We find that 
objects with CLASS$_{-}$STAR $\gtrsim$ 0.95 are typically stellar-like
based upon their locus in the magnitude vs. peak surface brightness diagram
and a visual inspection of their isophotal shapes.
(CLASS$_{-}$STAR $\rightarrow$ 1.0 are stars and
CLASS$_{-}$STAR $\rightarrow$ 0.0 are galaxies). Because our clusters
typically cover less than 2\% of the total area of an image,
we use individual images to statistically subtract
background galaxies.  By using our own images
rather than published background counts, we sidestep such issues as
differences in completeness and variations in counts due to 
projected large-scale structures.

\begin{figure*}
\figurenum{4}
\plotone{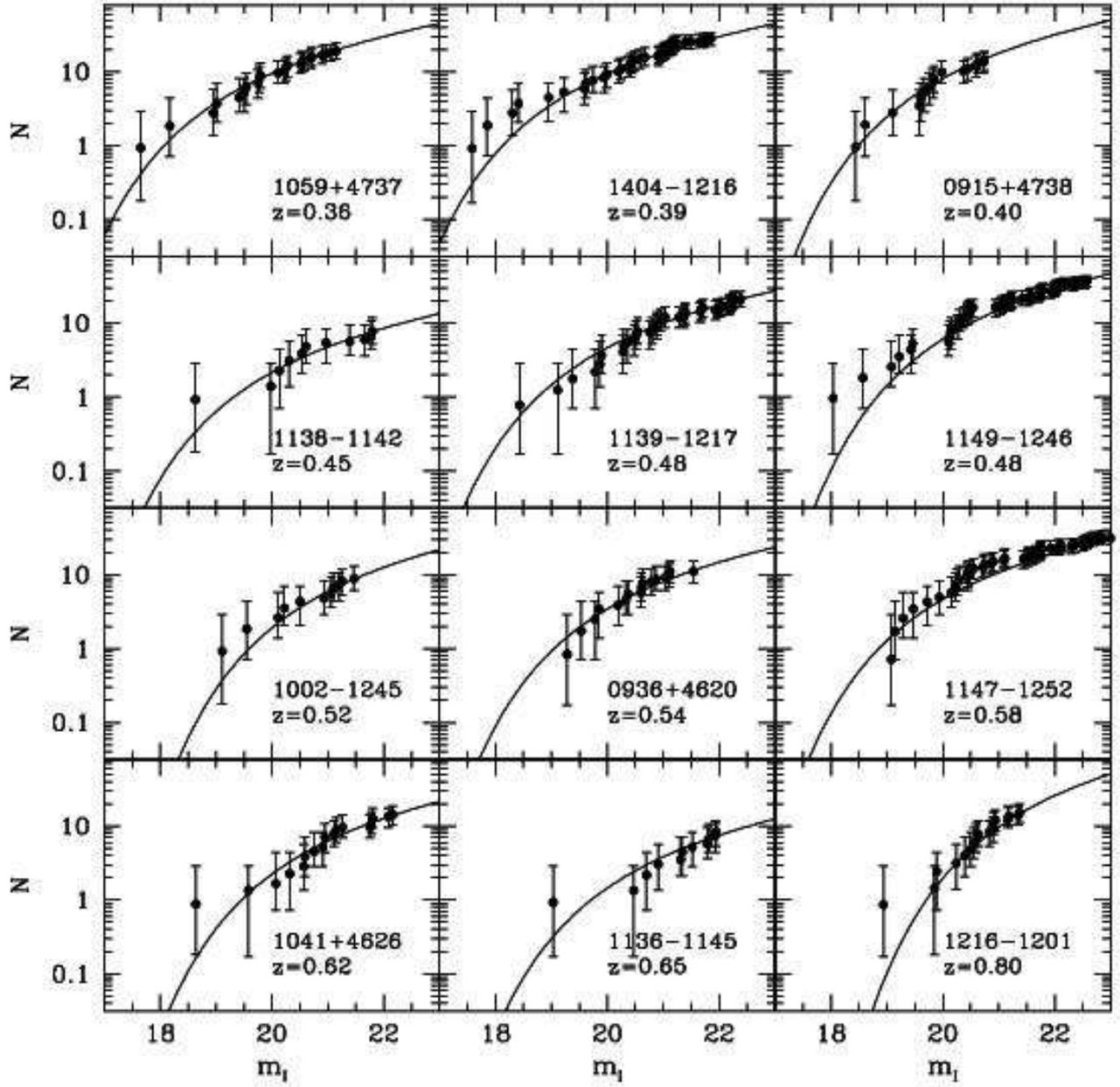}
\protect\figcaption{The cumulative number of galaxies as a function of
magnitude for our spectroscopically confirmed clusters with $I$-band
data. Overlaid is the best fitting Schechter function (\textit{solid
line}).}
\label{fig:lfcumm}
\end{figure*}

We vary \mstar and search for the
minimum $\chi^{2}$ to find the best fitting Schechter function.  We
apply this procedure using two different cumulative luminosity
distributions. The first distribution is the cumulative \textit{number} of
galaxies as a function of magnitude, while the second is the
cumulative \textit{luminosity} of galaxies as a function of magnitude.  The
first distribution weights the faint end more heavily
(there are more faint galaxies
and therefore necessarily more bins and more weight at the faint end)
and is consequently less affected by statistical variations. The
second distribution, on the other hand, weights the bright end more
heavily where the errors in our photometry are small. The fitting begins at a
magnitude corresponding to the brightest galaxy remaining after
background subtraction (presumably the BCG). Defining the faint end of
the fit is more difficult because each cluster has a different
completeness limit. Rather than determining the completeness of each
image, we exploit the fact that the fits to our two luminosity
functions are separately more sensitive to the bright and faint ends. Therefore,
changes in the faint end limiting magnitude affect one much more strongly
than the other. The faint end limit is initially
set to be m$_{I} = 23$ (none of our data is complete to this
magnitude). This limit results in a severe underestimation of the redshift
based on fits to the cumulative number of galaxies. Fits to the
cumulative luminosity of galaxies also underestimate the redshift,
but to a far lesser degree. We then decrease the faint end limiting magnitude until the
two redshift estimates agree to better than 15\%. The
average of these two estimates is taken to be the redshift, $z_{lum}$,
of that cluster. In Figure~\ref{fig:lfcumm}  we plot the cumulative number of
galaxies as a function of magnitude for our 12 spectroscopically
confirmed clusters with $I$-band data. The solid line is the best
fitting Schechter function corresponding to $z_{lum}$.

\begin{figure*}
\figurenum{5}
\plotone{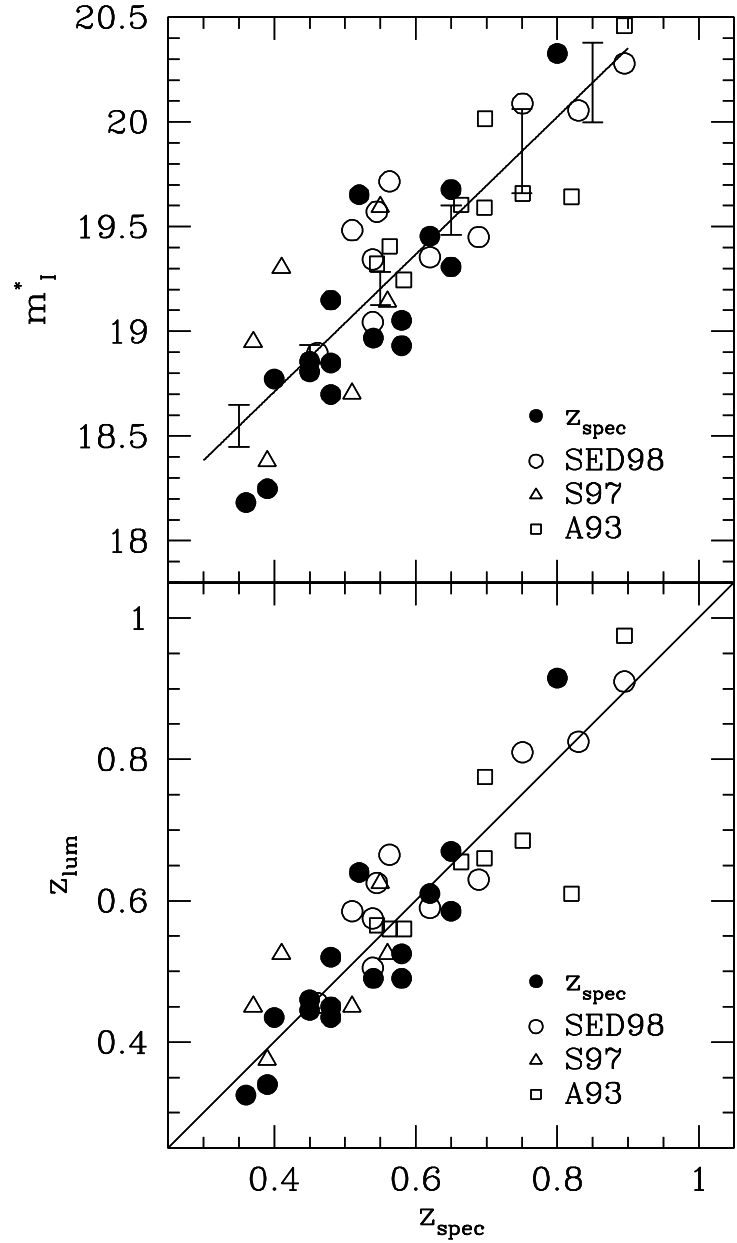}
\protect\figcaption{\textit{upper panel:} The derived value of \mstar vs. $z_{spec}$ for clusters from our survey, A93, S97, and SED98.  The line is a linear
fit to the data.  The
error bars are $\sigma_{mean}$ for clusters binned in $\Delta z = 0.1$,
where the mean is taken to be the value of the fit corresponding to
the center of the redshift bin.  \textit{lower
panel:} Comparison of $z_{lum}$ vs. $z_{spec}$.  The line is the
expected 1:1 correlation between the two redshifts indicators and has an
rms dispersion of 0.06.}
\label{fig:mstar}
\end{figure*}

We calibrate the \mstar - $z$ relation using
clusters with spectroscopic redshifts (upper panel of
Figure~\ref{fig:mstar}).  We parameterize the
relation with a linear function,
\mstar$ = {\rm C_{1}} + {\rm C_{2}}z $, with best
fitting parameters C$_{1} = 17.40\pm0.11$ and
C$_{2}=3.28\pm0.24$ (the rms about the fit is 0.28).  The plotted error bars represent 
$\sigma_{mean}$ of the data binned in
redshift intervals $\Delta z = 0.1$, where the mean is taken to be the
value of the fit corresponding to the center of the bin.  
The lower panel of Figure~\ref{fig:mstar} compares
$z_{lum}$ and $z_{spec}$.  Because the uncertainties in $z_{lum}$ are
much larger than those in $z_{spec}$, we adopt the rms scatter about
the line, 0.06, as the uncertainty in $z_{lum}$. 

\subsection{Color Envelope}

O'Connell (1988) defined the ``red envelope'' as the narrow locus of
red galaxies in the color-magnitude diagrams of intermediate
redshift clusters.  Subsequently the existence of the red envelope
has been confirmed for clusters extending to redshifts of
$z \sim $1 (A93; Stanford \etal 1995; Lubin 1996; SED98;
Kodama \etal 1998).  Smail \etal (1998) studied 10 x-ray luminous
clusters in the redshift range $z = 0.22-0.28$ and found a small
spread in the observed ($B - I$) colors ($\sim $0.04) for the brighter cluster ellipticals.  These
authors suggest that the location of the red envelope in clusters
would be an efficient, albeit somewhat crude, indicator of redshift
provided the relation could be calibrated with clusters spanning the
redshift range of interest.

We exploit the small scatter in the CM relation and use the location of
the red envelope in ($V - I$) as a redshift estimator, calibrating the relation
with a subset of our clusters that have spectroscopic redshifts.  The
location of the red envelope is not significantly affected by 
contamination from non-cluster galaxies, so we
include all galaxies within $350 h^{-1}$ kpc projected radius of the
cluster center.  For clusters without spectroscopy, we use our estimated redshift from the
luminosity function fitting, $z_{lum}$, to determine the projected
physical radius.  However, the resulting location of the color envelope is not
sensitive to relatively small changes in radius, thereby
providing an independent redshift indicator. 

\begin{figure*}
\figurenum{6}
\plotone{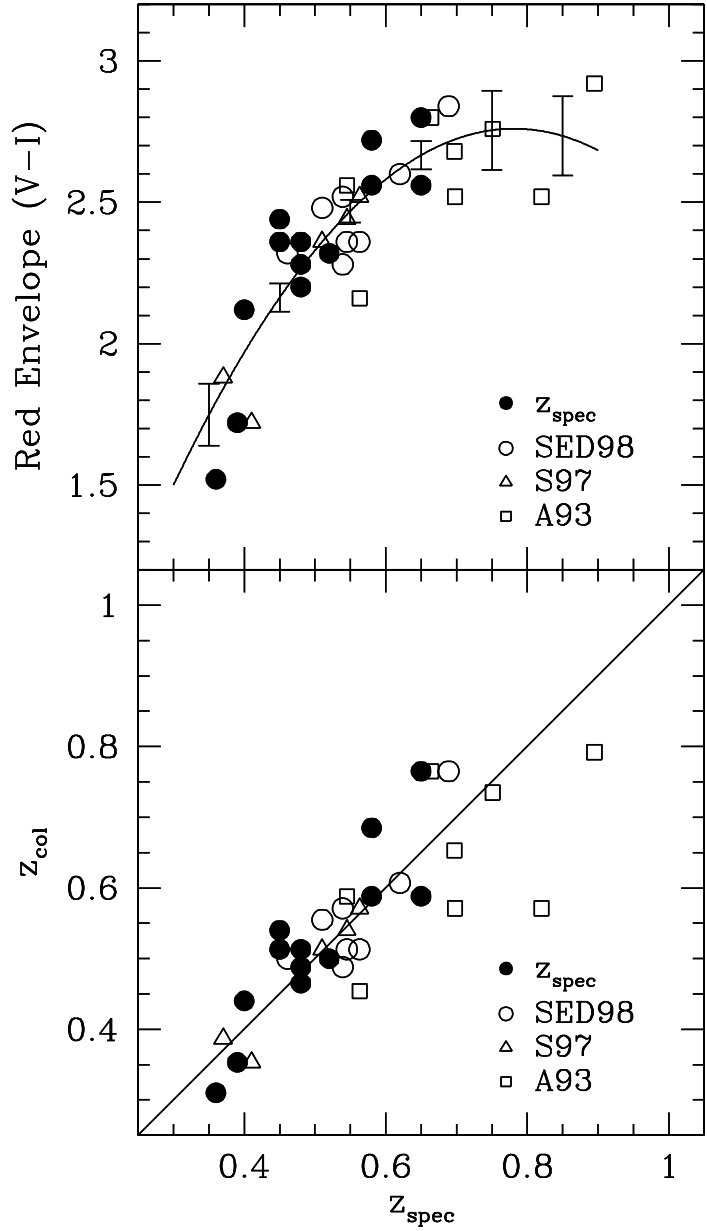}
\protect\figcaption{\textit{upper panel:} The location of the red envelope in ($V-I$) vs. $z_{spec}$ for clusters
from our survey, A93, S97, and SED98.  Error bars are $\sigma_{mean}$ for clusters binned in $\Delta z = 0.1$,
where the mean is taken to be the value of the fit corresponding to
the center of the redshift bin.  \textit{lower panel:} 
Comparison of $z_{spec}$ vs. $z_{col}$. The line is the expected
correlation of $z_{spec}$ and $z_{col}$ and has an rms dispersion of
0.07.}
\label{fig:redenv} 
\end{figure*}

\begin{figure*}
\figurenum{10}
\plotone{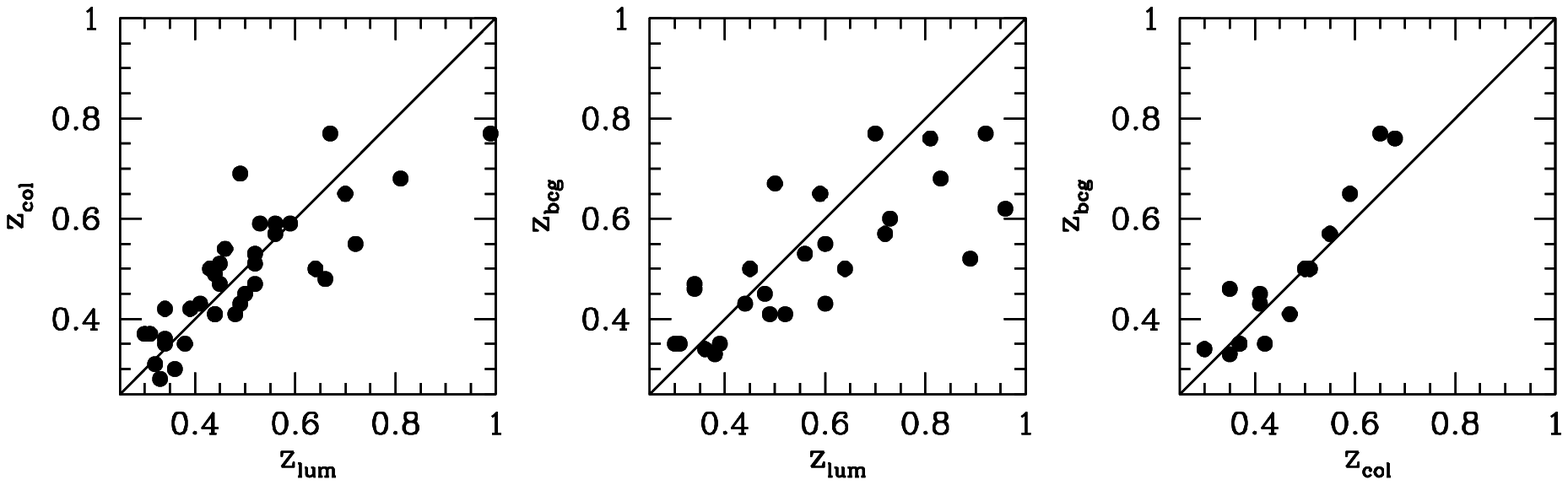}
\protect\figcaption{Comparison of the estimated redshifts derived using the three
photometric redshift estimation techniques, $z_{lum}$, $z_{col}$, and
$z_{BCG}$.}
\label{fig:photozs}
\end{figure*}

We define the location of the color envelope
to correspond to the maximum change in the number of galaxies
as a function of color.  This technique can be misled by
the presence of a tight clump of faint, blue galaxies in the CM
diagram.  To circumvent this
possibility, and because the red envelope is dominated by bright
ellipticals, we only consider galaxies with m$_{I} < 22$.  The red envelope position is determined by an automated routine and subsequently checked by a visual inspection of the
cluster's CM diagram. Figure~\ref{fig:redenv}, 
upper panel, shows the location of the red
envelope in ($V-I$) as a function of redshift for the subset
of our clusters that have spectroscopic redshifts, along with
clusters from A93, S97, and SED98.  We fit the empirical relation
with a parabolic function of the form
$(V-I) = $C$_{1} + $C$_{2}z + $C$_{3}z^{2}$, where the best
fitting parameters are C$_{1}=-0.56\pm0.53$, C$_{2}=8.50\pm1.82$, and
C$_{3}=-5.44\pm1.50$ (rms about the fit is 0.16).  
As for the \mstar vs. $z$ relation, 
the error bars represent $\sigma_{mean}$ of the
data binned in redshift intervals $\Delta z = 0.1$, where the mean is
taken to be the value of the fit corresponding to the center of the
bin. At high redshift ($z \gtrsim 0.6$) there are only small differences
in the $V-I$ colors of bright cluster ellipticals and the empirical
relation flattens. Therefore, clusters with $z_{col}\gtrsim0.6$ should
be regarded as lower limits to the true cluster redshift.
In the lower panel of Figure~\ref{fig:photozs} we plot
$z_{col}$ vs. $z_{spec}$. 
The uncertainties in $z_{spec}$ are
negligible compared to those in $z_{col}$ and therefore we set
$\sigma_{z_{col}}$ equal to the rms about the line,
0.07. However, because the bulk of the clusters used in this
determination have $z_{spec}\lesssim0.6$ where the empirical relation
is steep, this error estimate does not apply to high redshift clusters
whose value of $z_{col}$ is merely a lower limit. The small number of confirmed clusters at
high redshift does not allow us to estimate the errors associated with
the lower limiting values of $z_{col}$.

\begin{figure*}
\figurenum{7}
\plotone{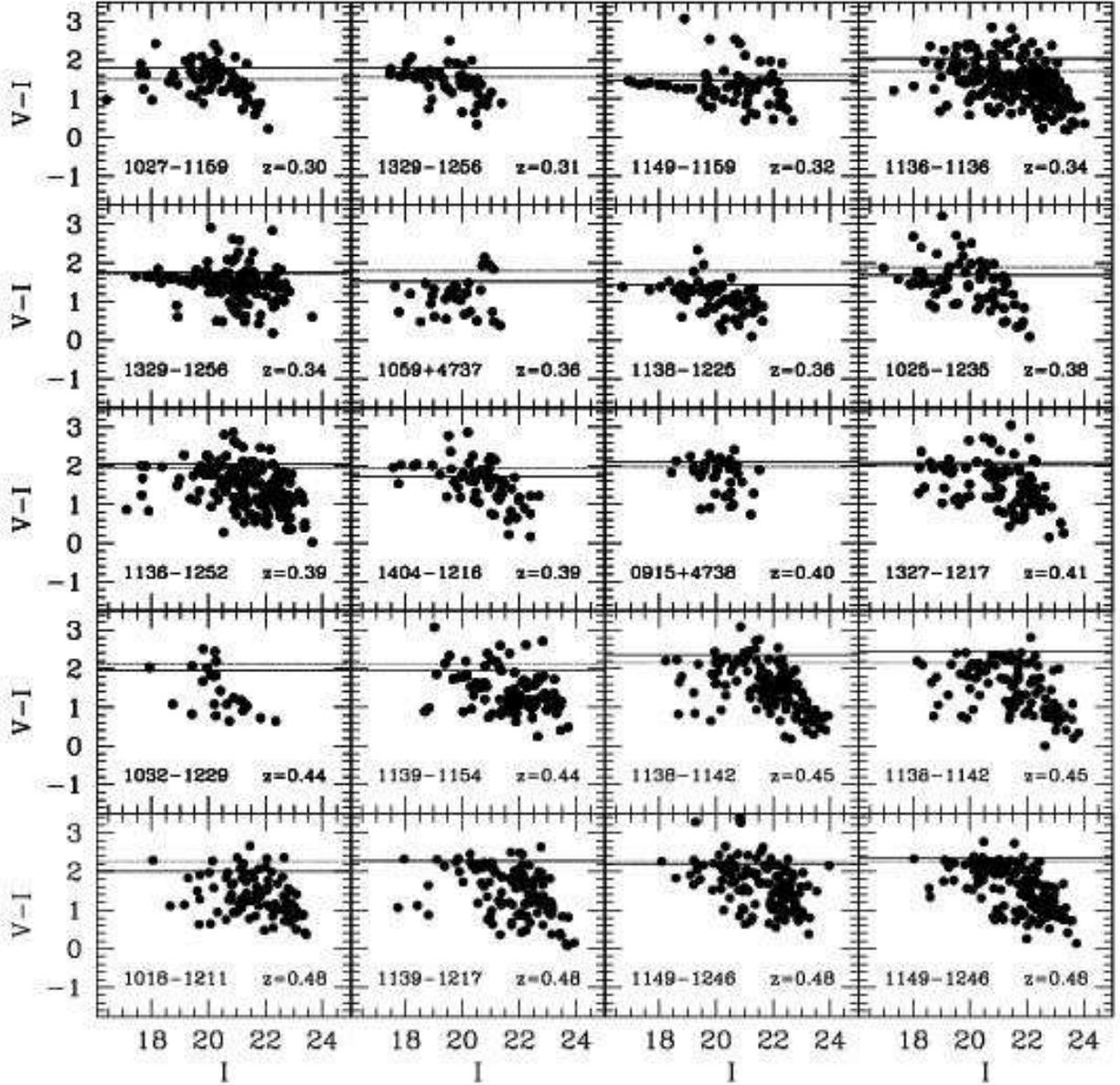}
\protect\figcaption{Optical color magnitude diagrams for our clusters, ordered in
increasing redshift ($z_{lum}$ is used if $z_{spec}$ does not exist).  Solid lines are the location of the red
envelope as determined by our automated procedure (see text for
details).  The dashed line corresponds to the expected location of the
red envelope based on the cluster's redshift and our $(V-I)$ vs. $z$ 
relation.}
\label{fig:cm}
\end{figure*}

Color magnitude diagrams for all candidates with follow-up
$V$ and $I$ data that we consider to be true clusters are
shown in Figure~\ref{fig:cm}.  These diagrams include all
galaxies within $350 h^{-1}$ kpc and are not background subtracted.
The clusters are shown in order of
increasing redshift according to $z_{lum}$, if $z_{spec}$ is
unavailable.
The solid line shows the location of the red envelope as determined by
the automated procedure described above.  The dashed line corresponds
to the expected location of the red envelope using our derived
$(V-I)$ vs. $z$ relation and the cluster's known or independently
estimated redshift.
For the majority of clusters, the agreement between the two
estimates is excellent.

\subsection{Classification of Cluster Candidates and Their Photometric
Redshifts} 

Without spectroscopy
of a number of galaxies per cluster field, the confirmation of
candidate clusters is indirect.  We use the
following criteria as guidelines in the classification process:
1) the appearance of the background subtracted $I-$band luminosity function,
in particular whether there is a well populated and defined bright end of the
luminosity function,
2) the presence of a prominent red envelope in the ($V-I$) CM diagrams, and
3) a clear peak in the surface density of galaxies, in particular a high degree of spatial clustering of red galaxies on the image.  Any 
cluster candidate that strongly satisfies any of these three criteria is 
considered a viable cluster. However, we do not require successful 
candidates to satisfy all three criteria for two reasons. First, we do not 
have the necessary 
data to evaluate every cluster equivalently. Second,
we have confirmed clusters that do not strongly satisfy all three criteria 
(for example, high-z clusters generally do not have prominent red 
envelopes due to the combination of cosmological shifting of light out of the
$V$-band and our modest exposure times). 

\begin{figure*}
\figurenum{8}
\plotone{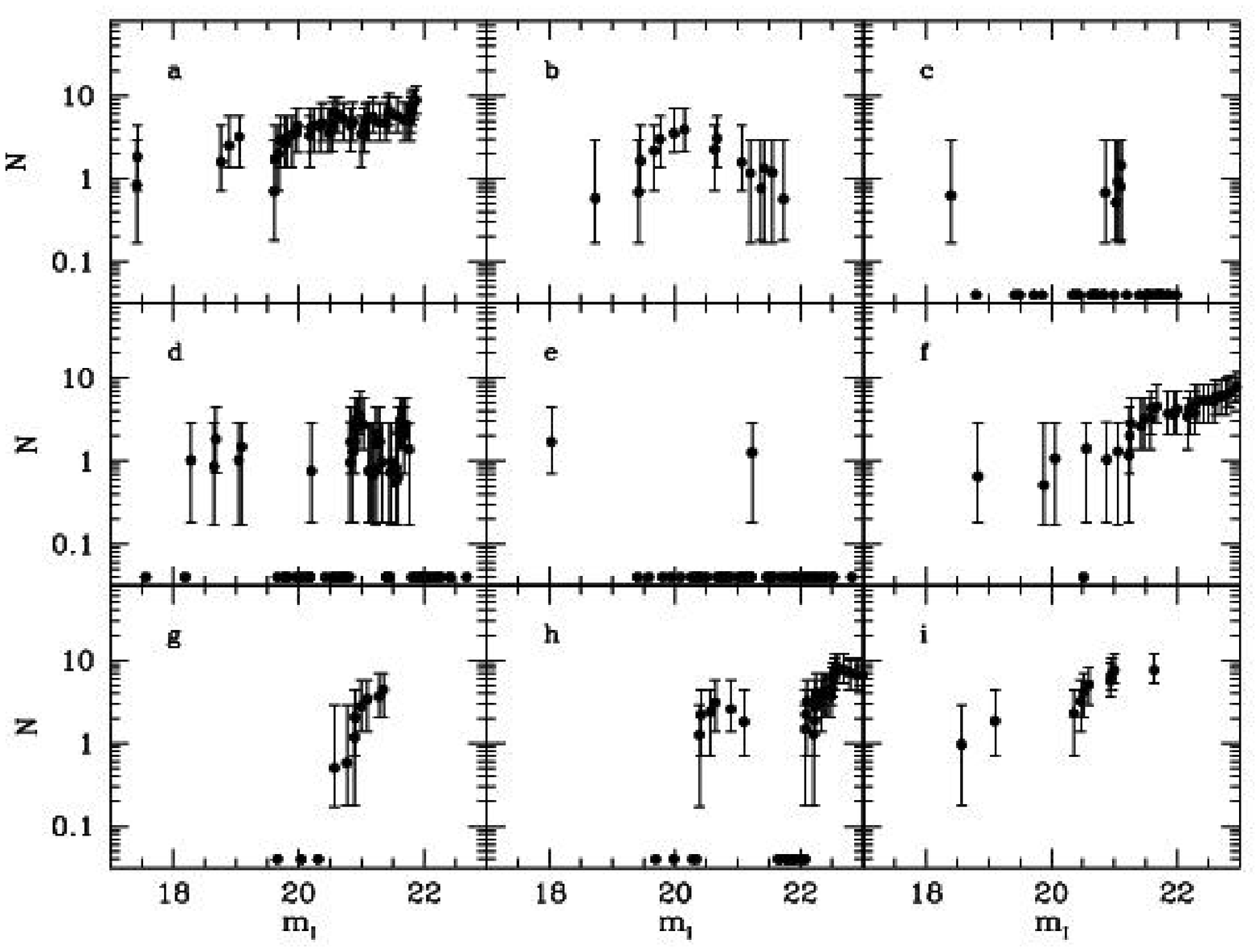}
\protect\figcaption{Cumulative luminosity functions for random fields on our
images (\textit{panels a-f}) and non-cluster detections
(\textit{panels g-i}). See text for details.}
\label{fig:lfrandom}
\end{figure*}

\begin{figure*}
\figurenum{9}
\plotone{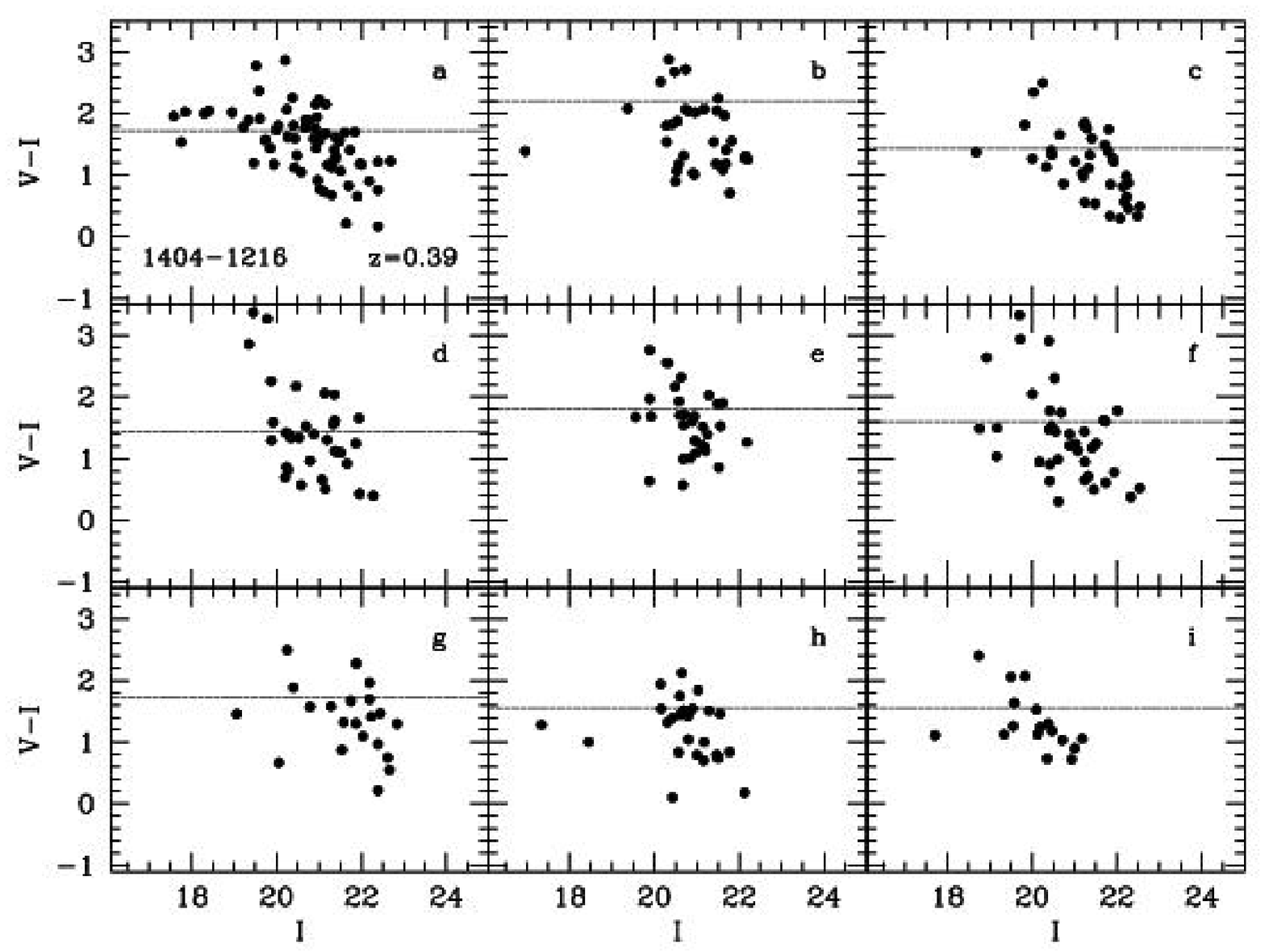}
\protect\figcaption{Color-magnitude diagrams for random fields on our images
(\textit{panels b-f}) and non-cluster detections (\textit{panels
g-i}). We reproduce the CM diagram of cluster 1404$-$1216, a
spectroscopically confirmed cluster at $z = 0.39$, for reference
(\textit{panel a}). The location of the
maximum change in the number of galaxies as a function of redshift as
determined by our automated routine is indicated by solid
lines.}
\label{fig:cmrandom}
\end{figure*}

Due to the heterogeneous nature of the data set and the intermediate
stage of our cluster selection algorithms, we acknowledge that our
classification procedure is subjective. However, we
demonstrate that random fields and non-cluster detections (LSBs and
spurious detections) do not yield well-defined luminosity functions nor
prominent red envelopes. In Figure~\ref{fig:lfrandom} (\textit{panels a-f}) we
plot luminosity functions of random fields on our images. Negative
values are arbitrarily assigned a value of 0.04. The random
centers are chosen to lie entirely outside the cluster region. The
random cluster radius is determined from surface density profiles of
the random fields in the same manner as for clusters. Because these
fields do not contain sharply peaked overdensities of galaxies, their
fitted radii are generally much larger than our typical cluster
radii. Figure~\ref{fig:lfrandom} clearly demonstrates that random fields do not
produce sensible luminosity functions. Next we plot the luminosity functions of three candidates
classified as either an LSB galaxy (\textit{panel g}), a spurious detection
(\textit{panel h}), and a doubtful cluster (\textit{panel i}). These
three objects are typical of our non-cluster detections in that their
luminosity functions are either very
steep (leading to $z_{lum} \gtrsim 1$), noisy, or flattened
(leading to $z_{lum} \lesssim 0.25$). The third object is a good illustration of
a borderline case (Figure~\ref{fig:lfrandom}, \textit{panel i}). The luminosity
function is too uncertain to use as the sole criteria for
classification as a cluster. Therefore, because this object lacks both color information and a
strong clustering of galaxies near the LSB detection peak, we exclude
it from our current analysis. Deeper follow-up
imaging of this candidate is necessary to finalize its classification.
Next, we examine the CM diagrams of random fields and non-cluster
detections. For reference, we reproduce the CM diagram of cl1404$-$1216, a spectroscopically
confirmed cluster at $z = 0.39$ (Figure~\ref{fig:cmrandom}, \textit{panel a}). We
choose 5 random centers on the image that lie entirely outside the
cluster region ($r > 700~h^{-1}$ kpc) and consider all galaxies within
$350 h^{-1}$ kpc of the random center to generate CM diagrams of
random fields (Figure~\ref{fig:cmrandom}, \textit{panels b-f}). The solid
line shows the maximum change in the number of
galaxies as a function of color as determined by our automated
procedure. The random fields do not show any coherent structure that
might be mistakenly identified as a red envelope. Similarly there is no indication of the presence of a red envelope in the CM diagrams for three rejected cluster candidates (Figure~\ref{fig:cmrandom},
\textit{panels g-i}).

Based upon our three criteria, Table 2 lists the 65
candidates that we classified as clusters, along with $z_{spec}$,
$z_{lum}$, $z_{col}$, $z_{BCG}$, m$^{*}_{I}$, the location of the red envelope in
($V - I$) and ($I - $\Kp).  Only the results using these clusters are presented in the
photometric redshift calibrations previously discussed.  Of our 65
clusters, 58 (92\%) have $z_{lum}$, 36 (57\%) have $z_{BCG}$, 30
(48\%) have $z_{col}$, and 17 (27\%) have $z_{spec}$.
Figure~\ref{fig:photozs} compares the estimated redshifts
derived using the three photometric redshift estimation techniques:
$z_{lum}$ vs. $z_{col}$ (\textit{left panel}),
$z_{lum}$ vs. $z_{BCG}$ (\textit{center panel}), and
$z_{col}$ vs. $z_{BCG}$ (\textit{right panel}).  All three techniques
correlate with $>$99\% confidence according to the Spearman rank test
and have rms scatters about the 1:1 lines that are in
agreement with the errors adopted for each photometric redshift
estimator.  Both of these factors convince us that these
objects are bonafide clusters. Of the 112 cluster candidates for which
we obtained follow-up imaging,
we rejected 49 objects (44\%). This relative success or failure rate
is not indicative of the final cluster catalog, however, because we specifically
tested the limits of our previous cluster selection algorithms by
selecting some marginal candidates.

\section{Discussion}

We now reverse roles and use our various photometric measures to 
study cluster galaxy evolution. In comparison to other recent studies
with superior data on individual clusters (eg. S97, SED98), we cannot
study any of our clusters in similar detail. However, we provide an
independent set of clusters and a significant enlargement of the sample
size with the combined dataset. For example, in the study of the evolution 
of \Mstar, we increase
the sample of clusters at these redshifts by 200\% (see \S4.2). 
Because we are searching for general trends,
it is arguable that more clusters with less precise data may be a superior 
sample than fewer clusters with more precise data.  Clearly, the two
types of studies are complementary.

In such an endeavor, one is
always uncertain whether the same class of object is being compared 
between high and low redshifts. We take complimentary approaches in this
discussion to root out potential difficulties.
We begin by examining the integrated population as characterized by
the luminosity evolution of \Mstar as a function of redshift.  Next we
focus on the galaxies that occupy the very narrow
locus in color in the cluster CM diagram characteristic of elliptical
galaxies. While we cannot guarantee
that galaxies do not move in and out of any one class of galaxies (in
fact we will later suggest that they do), we
can at least describe how that class evolves. Measurements of well-specified populations at the very least provide clear targets for simulations.
Finally, because our
survey is biased towards more massive clusters at high redshift, we search for
correlations between these empirically measured cluster galaxy properties
and cluster mass.

\subsection{Luminosity Evolution}

\begin{figure*}
\figurenum{11}
\plotone{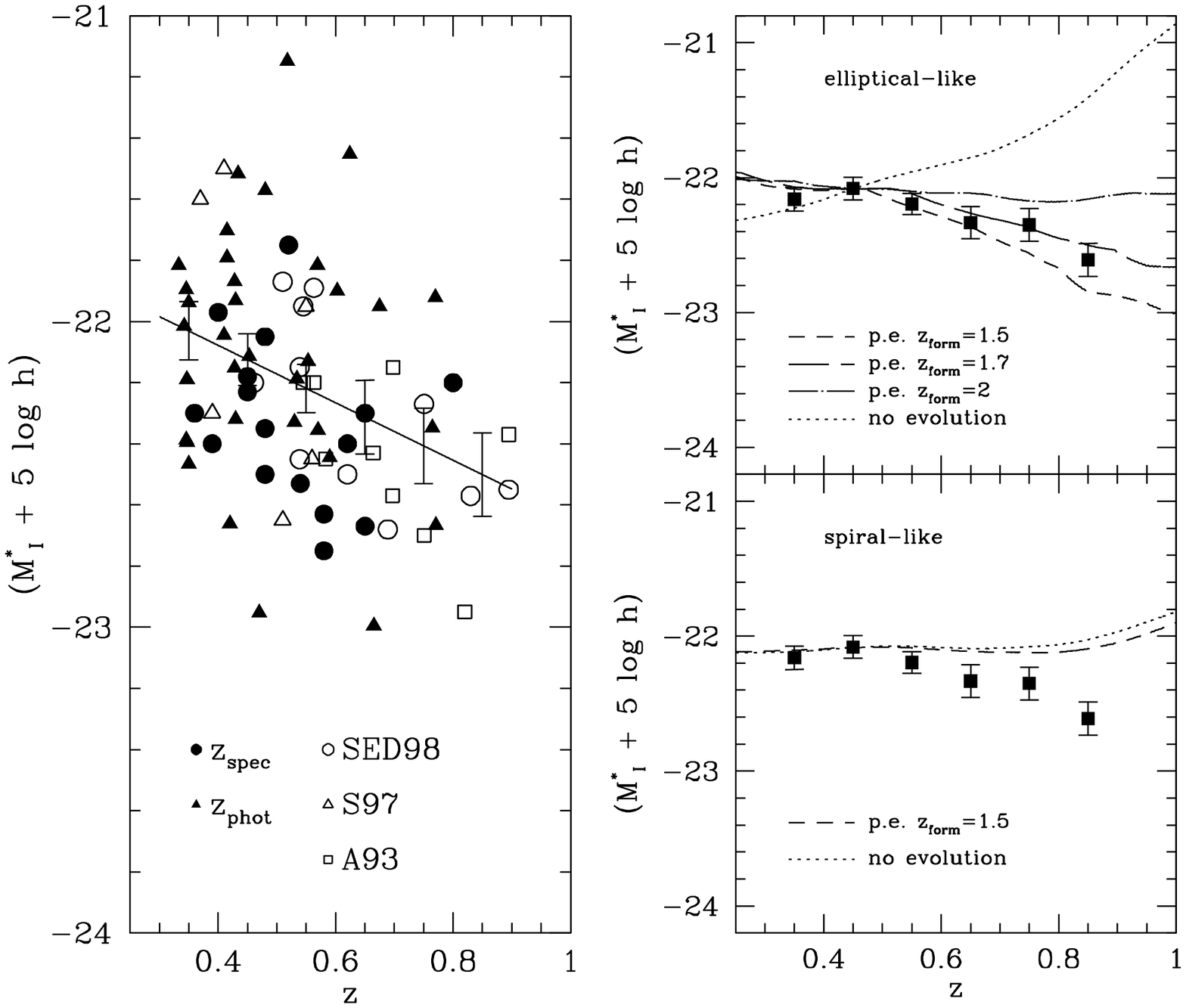}
\protect\figcaption{\textit{left panel:} The redshift
dependence of \Mstar vs. $z$ (in
the observed $I$ frame) for our clusters with spectroscopic and photometric
redshifts.  Also included are clusters from A93, S97, and SED98.
Error bars are $\sigma_{mean}$ for clusters binned in $\Delta z = 0.1$,
where the mean is taken to be the value of the fit corresponding to
the center of the redshift bin.  \textit{right panel:} 
Comparison of our observed values of \Mstar with the
expected luminosity evolution of both an elliptical (\textit{upper})
and spiral (\textit{lower}) galaxy using spectral
synthesis models from Bruzual \& Charlot (1993). The data are binned
in $\Delta z = $0.1 and have errors bars which are $\sigma_{mean}$ of
the binned data.  Both no evolution and
passive evolution models are considered.  The models are normalized at
$z = $0.45 which corresponds to the center of the redshift bin with the
greatest number of spectroscopically confirmed clusters.}
\label{fig:mste}
\end{figure*}

One way to characterize luminosity evolution in
cluster galaxies is to parameterize changes in the value of \Mstar as a
function of redshift.  The value of m$^{*}_{I}$ is a
by-product of our luminosity function photometric redshift estimation
technique. If an independent determination of redshift is
available, \mstar and  $z$  can be used to constrain evolution in
\Mstar.  The left panel of Figure~\ref{fig:mste} shows the derived
value of \Mstar both for clusters with spectroscopic redshifts
and those that have
either $z_{BCG}$ or $z_{col}$.  
For clusters with both $z_{BCG}$ and 
$z_{col}$, we use
$z_{BCG}$ because the associated uncertainties are smaller and because 
$z_{BCG}$ is less
likely than color to be directly tied to the
overall evolution of the bright end of the galaxy luminosity
function. There is a slight tendency for 
our spectroscopic clusters to have a brighter \Mstar than our non-spectroscopic
sample, although we
find that the difference 
is statistically insignificant.
The galaxy magnitudes are corrected for
Galactic extinction, but are not K-corrected.
Although there is considerable scatter in \Mstar at any $z$, 
a correlation exists
between \Mstar and $z$ at the $>$99\% confidence 
level according to a Spearman rank correlation test.
We parameterize the full data set with a linear function of the form
(\Mstar$+5 \log h) = {\rm C_{1}}+{\rm C_{2}}z$, with best fit
parameters of C$_{1}=-21.74\pm0.12$ and C$_{2}=-0.88\pm0.24$ (rms
about the fit is 0.34). The 
error bars in Figure~\ref{fig:mste} represent
$\sigma_{mean}$ for clusters binned in $\Delta z=0.1$,
where the mean is taken to be the value of the fit corresponding to
the center of the redshift bin.  
This parameterization is valid over 0.3$\leq z \leq$0.9, the redshift
range probed by our sample, and for $\Omega_{0} = 0.2 (\Omega_\Lambda = 0)$.
We conclude that the luminosity in the observed $I$ frame
of a characteristic cluster galaxy is
$0.53\pm0.28+5 \log h$ mag brighter at $z = 0.9$ than at $z = 0.3$.

\begin{figure*}
\figurenum{12}
\plotone{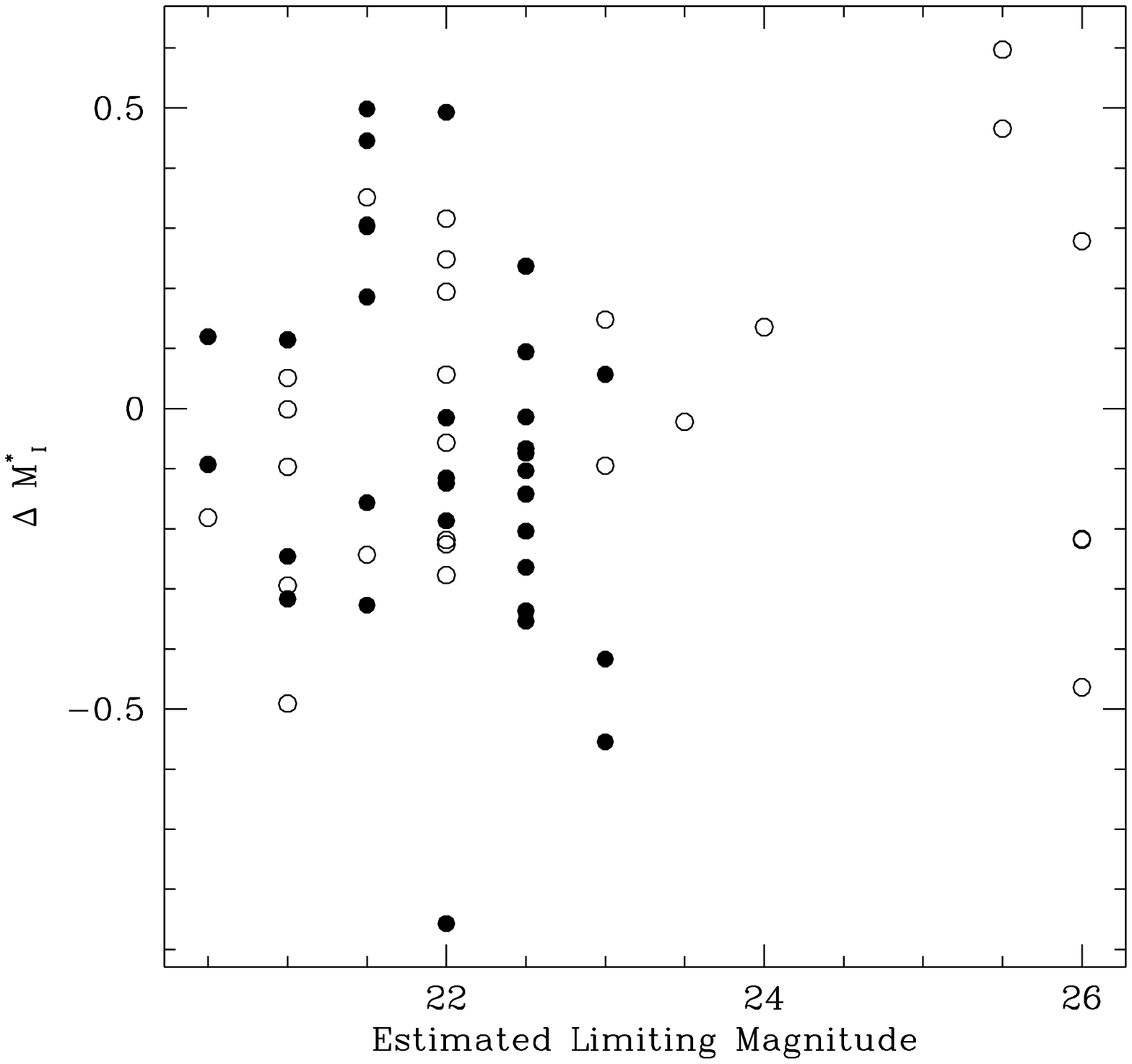}
\protect\figcaption{Comparison between the observational limiting magnitude
and the measured m$_*$.}
\label{fig:maglim}
\end{figure*}

We test for possible systematic errors in a variety of ways.
First, this result is consistent with that obtained using
only clusters with spectroscopic redshifts (for which \Mstar is
$0.59\pm0.17+5 \log h$ mag brighter over the same redshift range),
which argues against the influence of potential problems with
the photo-z measurements. Second, the result is consistent with
that obtained using only the literature clusters, which again
argues against the influence of problems with either our photometry,
the fitting of m$^*_I$, or the photo-z's. Lastly, we find that
our measurement of m$^*_I$ is independent of the limiting magnitude of
the observations, where we estimate the completeness limit of each field
using the color-magnitude diagrams and identifying the $I$ magnitude
at which the red sequence is lost. There is no evidence
that the magnitude limit affects our determination of m$^*$ 
(Figure~\ref{fig:maglim}).

How much of this apparent evolution is due to K-corrections and
how does our observed evolution in \Mstar compare to theoretical
predictions?  A direct comparison requires detailed modeling of the
formation and evolution of the cluster galaxy population as a whole
and is beyond the scope of this paper.  We take a simpler
approach and compare the evolution of \Mstar to the
expected luminosity evolution of both a spiral and an elliptical
galaxy.  We caution that due to the various complications involved in
the interpretation of evolution of the luminosity function,
principally that we are studying the full wide range of morphological
classes present at these magnitudes, 
any conclusions drawn must be viewed as merely suggestive.  In 
particular, the error bars do not include uncertainties in the relative
normalization of the models and assume that the scatter in the measurements
is Gaussian about a single true value.

The right panels of Figure~\ref{fig:mste} compare our observations 
of \Mstar vs. $z$
to spectral synthesis model predictions obtained using the GISSEL96 code (Bruzual
\& Charlot 1993).  Because parameterizing data involves some
arbitrariness in the selection of the functional form, we choose to
compare directly to the data, binned in redshift intervals of $\Delta
z = 0.1$.  The associated error bars are $\sigma_{mean}$ of the binned
data.  We consider two types of star formation histories: 1)
an initial 10$^{7}$ Gyr burst of star formation (elliptical-like;
\textit{upper panel}),
and 2) an exponentially declining star formation rate such that
$\mu$ is the fraction of the galaxy's mass converted
into stars and is given by $1-e^{-1/\tau}$, with $\mu=$0.1 and $\tau$ in Gyr 
(spiral-like; \textit{lower panel}).
All models have Salpeter initial mass functions for masses
between 0.1$\msun$ and 100$\msun$.  The models are normalized to the
data at $z =$0.45, which corresponds to the center of the redshift bin
that has the greatest number of spectroscopically confirmed 
clusters.  The models are computed in the observer's frame and thus can be
directly compared to our observations.  

The most immediate conclusion is that no model using a spiral galaxy
SED fits the data.  No evolution and passive evolution models predict
roughly the same luminosity evolution for a spiral, regardless of
$z_{form}$, and therefore only two representative cases are plotted.
The model spiral galaxy's luminosity remains roughly constant over the
redshift range $0.3\leq z \leq1$ and is inconsistent with the
observed brightening of \Mstar. Turning to
the elliptical galaxy evolution models, we find that the no evolution
model is ruled out.  Also, passive
evolution models with $z_{form} > 2$ are poor fits to the data.
In these models, a \Mstar galaxy formed at a sufficiently early epoch that
appreciable star formation ceased well before $z\sim$1, so the galaxy 
is not evolving sufficiently over the redshift range probed by our sample to
counter the K-correction.  In both of these cases, \Mstar actually
dims with increasing redshift due to the cosmological shifting of
light out of our observed passband.  Our observations of \Mstar agree
best with passive evolution models with a relatively low $z_{form}
\sim 1.7$.  In these models, a \Mstar galaxy underwent its burst of star
formation rather recently, and we are simply tracking the decline in
luminosity due to the passive aging of the stellar populations. 
These models do not include mergers or accretion, so we only
infer from this result that the luminous galaxies in
these clusters are somewhat brighter than expected hat they had formed
at high redshifts and since been quiescent.

In contrast to some earlier studies that focus on the reddest optical
bands or even extend into the near-IR wavelengths (e.g. de Propris \etal 1999) to avoid the difficulty in modeling the rest-UV spectral energy
distribution of galaxies, or that use different passbands for
different cluster redshifts to minimize K-corrections (e.g. SED98),
we have used primarily $V$ and $I$ (and \Kp for a smaller fraction
of clusters) in the observed frame. We consider the use of relatively
blue rest-frame optical data an important complement to existing
data. Before comparing our conclusions to those of previous
studies, we discuss whether our approach compromises our study. The
disadvantage of using $V$ and $I$ is that it places greater demands on the
models because the rest-frame UV must be modeled well (observed $V$
corresponds to rest-frame 2900\AA\ at $z = 0.9$). The advantage is that
it probes the spectral range that is most sensitive to recent star
formation. The models may be more robust at redder wavelengths, but
the effect of recent star formation is more subtle. If our results
agree with previous studies, then we can conclude that the
theoretical models are working in the $U$-band and that the previous
conclusions regarding the evolution of cluster galaxies hold.  If our
results disagree with previous studies, then we can conclude either 1)
that the models may have problems in the $U$-band, 2) that our clusters
are different than those studied previously, 3) that evolution beyond
simple passive evolution is fairly subtle and only detectable at bluer
wavelengths, or 4) that there are systematic errors in either our data
or previous data (or both).

Our findings are in qualitative agreement with previous studies of
luminosity evolution.  Using WFPC2 images of Cl0939 ($z = 0.41$) and
ground-based $R$-band images of Coma, Barrientos \etal (1996) measured
the luminosity-effective radius relation for bright ellipticals and
determined that passive evolution models with $z_{form}\sim$1.2
provided the best fit to their data.  S97 measured composite
luminosity functions for early and late-type galaxies in 10 clusters
at 0.37 $\leq z \leq$ 0.56 from WFPC2 images.  The ellipticals show
a modest, but significant, brightening of M$^{*}_{V}$ consistent with
passive evolution and $z_{form}\sim3$.  In contrast, the late type
galaxies show no evidence for a brightening in M$^{*}_{V}$ over the
same redshift range.  Although our luminosity functions consist of the
entire range of morphological classes, we are limited to the bright
galaxies, which are presumably ellipticals.  While S97's luminosity
evolution best matches passive evolution with a slightly higher
$z_{form}$ than our observations, their sample only probes out to
$z = $0.56.  The upper right panel of Figure~\ref{fig:mste} demonstrates
that significant differences in passive evolution models do not appear
until $z > 0.6$ and thus we conclude that we are not in disagreement
with S97. Additionally, we are in agreement with de Propris \etal
(1999) who find that M$^{*}_{K}$ departs from no-evolution model
predictions at $z > 0.4$ and that this departure
is consistent with passive evolution with intermediate 
z$_{form}$. Again, because of many unquantifiable uncertainties we
do not highlight minor discrepancies between the studies, rather we
search for broad qualitative agreements or disagreements.

\subsection{Red Envelope Evolution}

\begin{figure*}
\figurenum{13}
\plotone{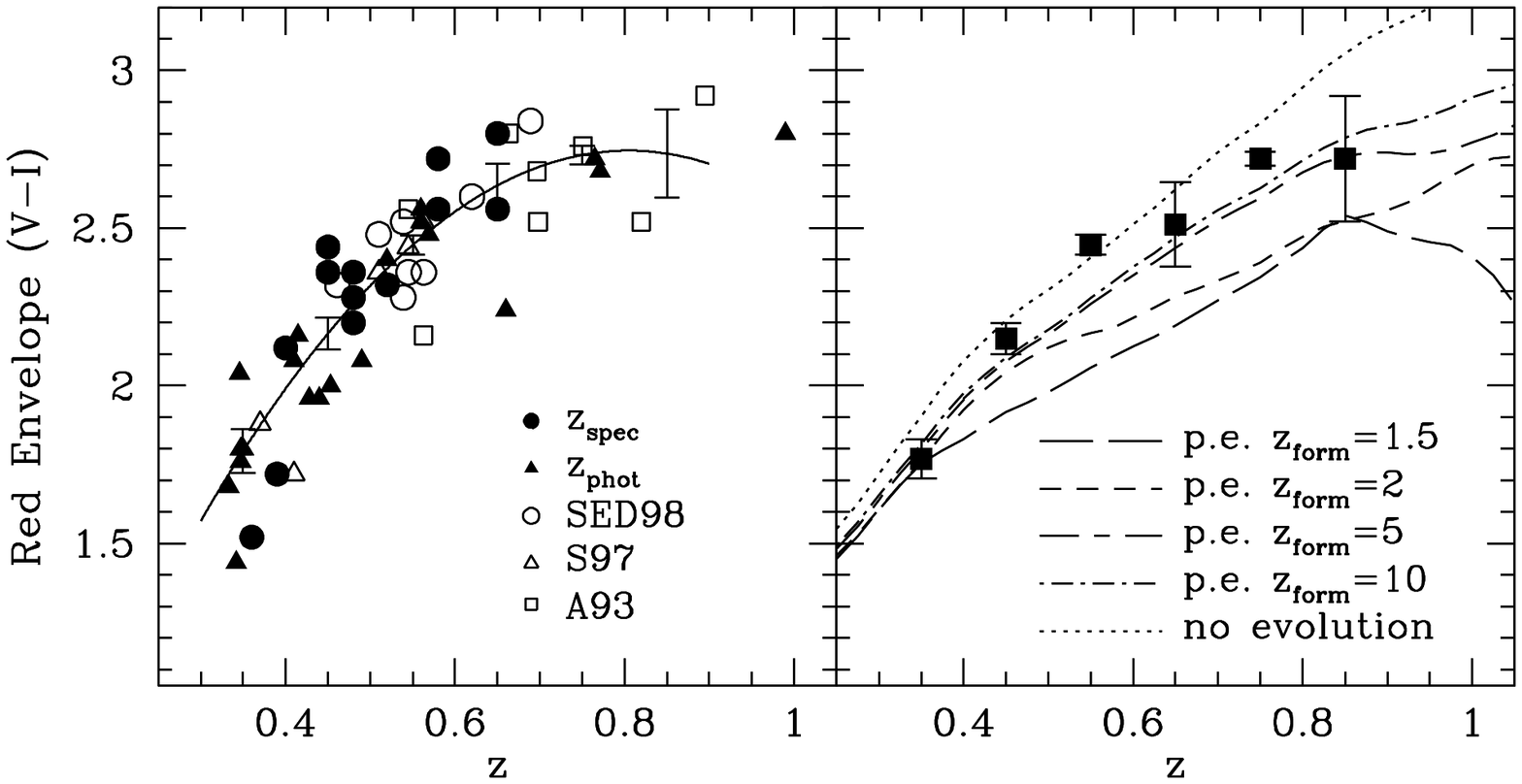}
\protect\figcaption{\textit{left panel:} The location of the red envelope in
$V-I$ vs. $z$ for our clusters with spectroscopic and photometric
redshifts.  Also included are clusters from A93, S97, and SED98.
Error bars are $\sigma_{mean}$ for clusters binned in $\Delta z = 0.1$,
where the mean is taken to be the value of the fit corresponding to
the center of the redshift bin.  \textit{right panel:} 
Comparison of the observed location of the red envelope in $V-I$ with the
expected color evolution of an elliptical galaxy using spectral
synthesis models from Bruzual \& Charlot (1993). The data are binned
in $\Delta z = 0.1$ and have errors bars which are $\sigma_{mean}$ of
the binned data.  Both no evolution and
passive evolution models are considered.}
\label{fig:redenv_vi}
\end{figure*}

\begin{figure*}
\figurenum{14}
\plotone{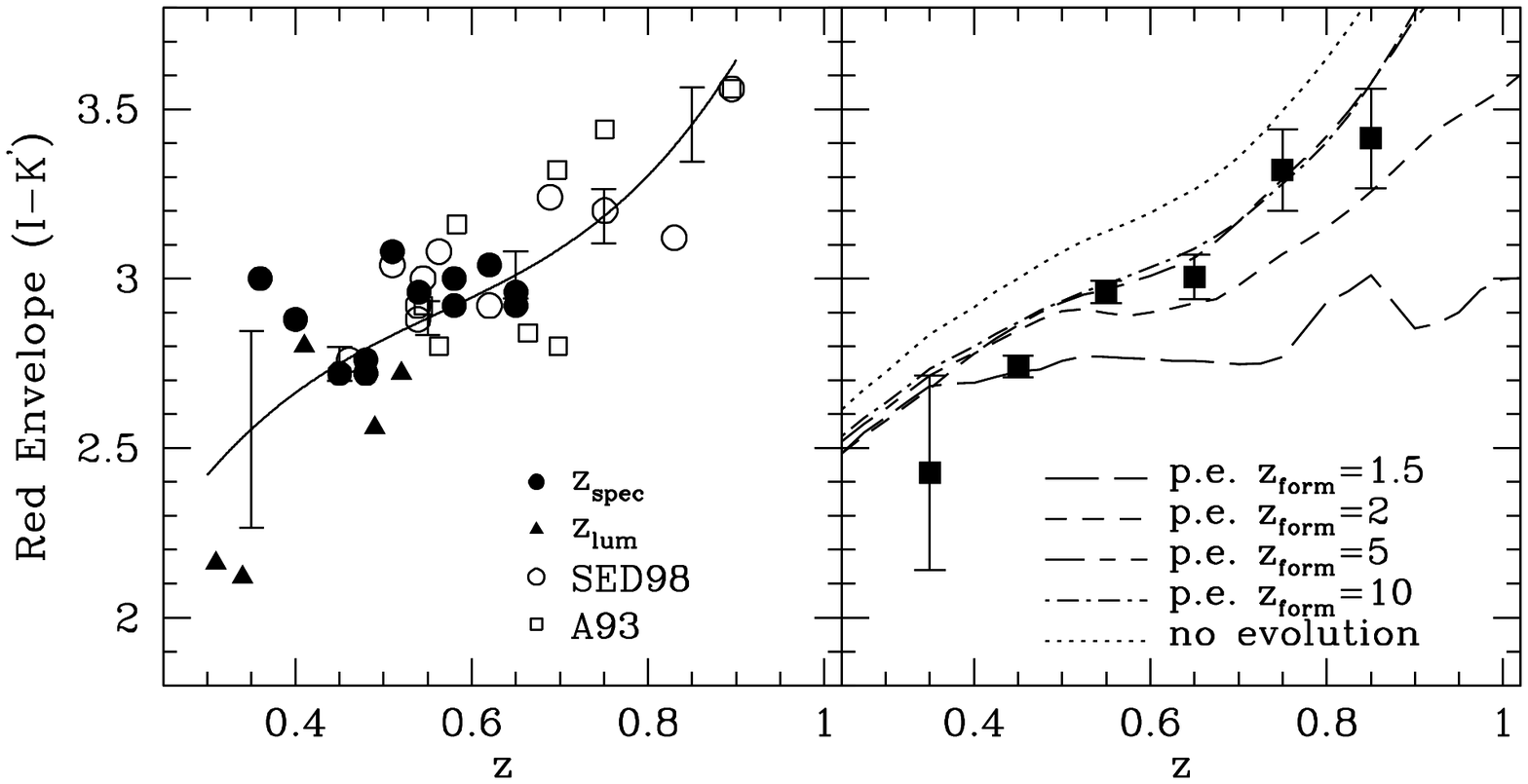}
\protect\figcaption{\textit{left panel:} The location of the red envelope in
$I-$\Kp vs. $z$ for our clusters with spectroscopic and photometric
redshifts.  Also included are clusters from A93 and SED98.
Error bars are $\sigma_{mean}$ for clusters binned in $\Delta z = 0.1$,
where the mean is taken to be the value of the fit corresponding to
the center of the redshift bin.  \textit{right panel:} 
Comparison of the observed location of the red envelope in $I-$\Kp with the
expected color evolution of an elliptical galaxy using spectral
synthesis models from Bruzual \& Charlot (1993). The data are binned
in $\Delta z = 0.1$ and have errors bars which are $\sigma_{mean}$ of
the binned data.  Both no evolution and
passive evolution models are considered.}
\label{fig:redenv_ik}
\end{figure*}

The previous approach analyzes the evolution in the combination of galaxy
populations, therefore the results are somewhat 
difficult to interpret. Alternatively,
we can select a single galaxy population and study its
evolution.
We investigate a group of galaxies that inhabit a
well-defined locus in the cluster CM diagram, the red envelope.
We plot the location of the red
envelope in $V-I$ versus redshift (Figure~\ref{fig:redenv_vi}  \textit{left
panel}) and in $I-$\Kp (Figure~\ref{fig:redenv_ik}, \textit{left panel}).  Again,
we use clusters with independent photometric redshifts (in this case
$z_{BCG}$ or $z_{lum}$) and clusters with spectroscopic redshifts to parameterize the relations.  For $V-I$ we fit the
observed relation with a parabolic function of the form $(V-I) =
$C$_{1}+ $C$_{2}z + $C$_{3}z^{2}$, where the best fitting parameters
are C$_{1}=-0.24\pm0.28$, C$_{2}=7.42\pm1.03$, and
C$_{3}=-4.61\pm0.91$ (the rms about the fit is 0.16).  For $I-$\Kp
the observations are best fit with a quadratic function of the form
$(I-$\Kp$) = $C$_{1} + $C$_{2}z + $C$_{3}z^{2} + $C$_{4}z^{3}$,
where the best fitting parameters are C$_{1}=0.66\pm0.65$, C$_{2}=9.50\pm3.72$,
C$_{3}=-14.72\pm7.01$ and C$_{4}=8.72\pm4.29$ (the rms about the fit
is 0.18).  Both
parameterizations are valid over $0.3 \leq z \leq 0.9$, the redshift
range of our clusters.  The error bars are $\sigma_{mean}$ for
clusters binned in $\Delta z = 0.1$, where the mean is taken to be the
value of the fit corresponding to the center of the redshift bin. We
summarize our empirically derived luminosity and color evolution
parameterizations 
in Table 3.

The brightest, reddest galaxies,
which are presumably the cluster's oldest ellipticals
populate the red envelope. We compare our
measured color evolution in ($V-I$) (Figure~\ref{fig:redenv_vi}, \textit{right
panel}) and ($I-$\Kp) (Figure~\ref{fig:redenv_ik}, \textit{right panel}) with
model predictions obtained using the GISSEL96 code (Bruzual \& Charlot 1993) for an elliptical
galaxy undergoing passive evolution or no evolution.  The parameters
of the models are the same as those given above.  When comparing
colors, in contrast to the magnitude comparisons of the previous
section, no normalization is necessary. Because there is a
certain amount of arbitrariness in parameterizing the data with a
function, here we simply bin the data in intervals of $\Delta z = 0.1$.
The error bars are $\sigma_{mean}$ of the binned data.  Turning first
to the evolution in ($V-I$), at low redshift the data lie intermediate
between no evolution and passive evolution models with a high
formation redshift,
$z_{form}>$5.  However, at high redshift where the differences in model
predictions increases, we find that the observations are inconsistent
with the no evolution prediction. Our data are most consistent with
models of elliptical galaxies with passive evolution and
$z_{form}\gtrsim$5.  Examining the color of the red envelope in
$I-$\Kp, we find that no evolution models are completely ruled out.
Such models predict colors that are too red compared to the location
of the red envelope at all $z$.  These data are most consistent
with a passive evolution model with a high epoch of formation,
$z_{form}\gtrsim5$. 

Similar conclusions  have been reached by previous authors, using both 
optical and infrared-optical colors.  RS measured
galaxy colors in the rest-frame Stromgren $\textit{u v b y}$ filters
and found a degree of blueing of their sample from $z \sim 0.4$ to 0.9 that
implies a redshift of formation of $\sim 10$.  From the PDCS 
using the WFPC2 bands $V_{4}$ (F555W) and $I_4$ (F785LP), Lubin \etal
(1996) find
the red envelope of their clusters is bluer in $V_{4}-I_{4}$ than no
evolution predictions, and implying passive evolution with $z_{form}>2$.
Using optical-infrared colors, both A93 and SED98 observe the blueing
trend compared to no evolution model predictions.  Their observations agree best
with models in which galaxies form at $ z \sim 2$ to 4 and 
passively evolve thereafter. We conclude that one can identify cluster
galaxies that appear to have formed the bulk of their stars at
$z \sim$ 5 by selecting galaxies on the red envelope.

\subsection{Cluster Selection Biases}

One concern that cannot yet be entirely addressed is that the
comparison across a large redshift range incorporates clusters with
different properties. We expect to detect only the richest, most
massive systems at high redshifts and a wider range of systems at low
redshifts. Because we do not have x-ray observations of our clusters,
we cannot yet address this potential bias directly. However, an
indirect cluster mass estimation technique indicates that 
indeed we only detect the most
massive systems at high redshift (see Gonzalez \etal 2001). Using 17
known clusters with x-ray data for which we have drift scans, we
find a correlation between the x-ray luminosity, L$_{x}$, and the peak
surface brightness of our cluster detection, $\Sigma$. Using this
correlation, we find that we detect clusters with a wide range in mass
(groups to rich clusters) at low to intermediate redshift. However, at
high redshift ($z \gtrsim 0.7$) we detect only the most massive systems
(e.g. we can detect MS 1054$-$03 with
L$_{x}\sim 9 \times 10^{44}$ ergs s$^{-1}$).

Whether the galaxy properties of these systems depends on the global
characteristics of the cluster is not well determined. Recent
work suggests that the color evolution of bright ellipticals and the
evolution of M$^{*}$ are not strongly
affected by cluster mass. Stanford \etal (1995) studied an
inhomogeneous sample of 19 clusters with $0 < z < 0.9$ drawn from a
variety of optical, x-ray, and radio-selected surveys that span a wide
range in cluster mass
($3.0 \times 10^{43}$ ergs s$^{-1} < $L$_{x} < 2.0 \times 10^{45}$ ergs s$^{-1}$).
They found the color evolution of the bright elliptical galaxies in
these clusters does not depend strongly on either L$_{x}$ or optical
richness. Similarly, Smail \etal (1998) found $\lesssim$2\% scatter in
the rest-frame UV-optical colors of the bright elliptical sequence in
10 x-ray luminous ROSAT clusters at $0.22 \leq z \leq 0.28$.

The
evolution of M$^{*}$ also does not seem dependent upon cluster
mass. Using a heterogeneous sample of 38 clusters at $0.1 < z < 1$,
de Propris \etal (1999) fit Schechter functions with a fixed
$\alpha = -0.9$ to their K-band data. They found no statistically
significant difference in the measured value of m$^{*}_{K}$ as a
function of $z$ when they
divided their sample based upon x-ray luminosity or optical
richness. A similar result is found in optical bands. Garilli,
Maccagni, \& Andreon (1999) studied a sample of 65 Abell and x-ray 
selected clusters at
$0.05 < z < 0.25$ in the Gunn $g$, $r$, and $i$ bands. They constructed
composite luminosity functions and found similar values of M$^{*}$ in
all three bands in both the Abell and x-ray selected samples and
also between subsamples divided by optical richness.

\begin{figure*}
\figurenum{15}
\plotone{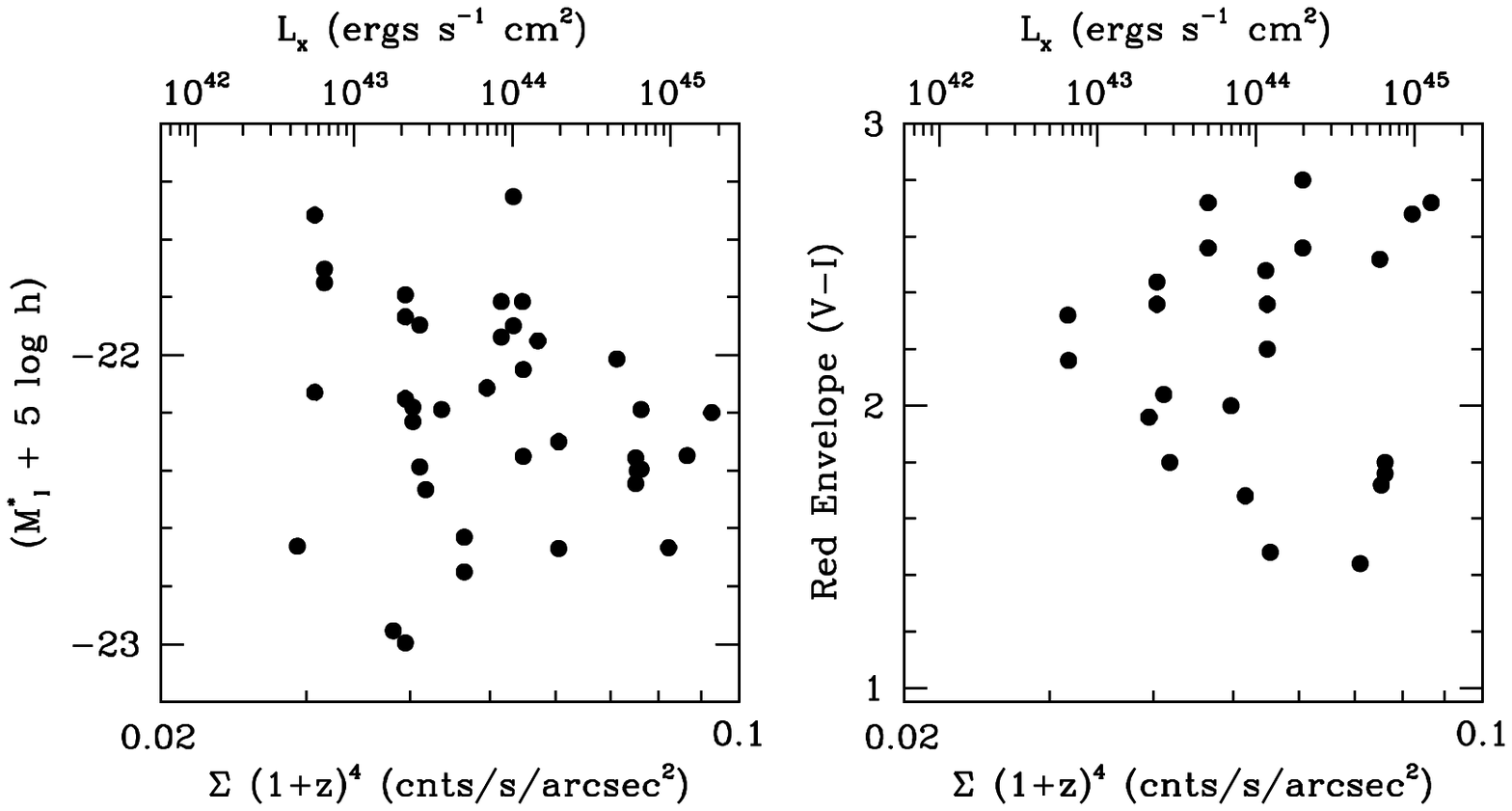}
\protect\figcaption{The peak surface brightness of the cluster detection,
$\Sigma (1 + z)^{4}$, compared
to (\Mstar$+5 \log h)$ (\textit{left panel}) and the location of the
red envelope in $(V-I)$ (\textit{right panel}). The values of L$_{x}$ are inferred from our
empirically determined correlation between $\Sigma (1 + z)^{4}$ and
L$_{x}$; they are not directly observed. Neither of the galaxy
properties are significantly correlated with
$\Sigma (1 + z)^{4}$ according to the Spearman rank test.}
\label{fig:sigma_color} 
\end{figure*}

We use our empirically derived correlation between the peak surface
brightness of the cluster detection, $\Sigma$, and x-ray luminosity,
L$_{x}$, and search for trends between cluster galaxy properties and
cluster mass. In Figure~\ref{fig:sigma_color}, we compare $\Sigma (1 + z)^{4}$
to (\Mstar$+5 \log h)$ (\textit{left panel}) and the location of the
red envelope in $(V-I)$ (\textit{right panel}). Neither of these quantities are significantly
correlated according to the Spearman rank test.
Although current work suggests that certain cluster galaxy properties are not
strongly dependent upon cluster mass, this 
issue merits further study and requires large cluster samples 
such as that developed here. Better mass estimates for these clusters are critical
in addressing this issue.

\subsection{Cluster Galaxy Evolution}

The various empirical results presented above, in conjunction with
published results, present a rather complex picture of cluster
galaxy evolution. First, the galaxies on the cluster 
E/S0 sequence appear to be evolving as elliptical galaxies with very large 
formation redshifts ($\gtrsim 5$). Second, 
the bulk of the luminous cluster galaxies (as described
by the luminosity function) appear to be evolving similarly to 
elliptical galaxies with low formation redshifts ($\lesssim 2$).
Are these two independent galaxy populations, or is there one
process that explains all of them?

It is evident that there are some cluster galaxies, those on the red E/S0
sequence, that consist of very old stellar populations. The comparison
of spectral 
synthesis models with the $V-I$ and $I-$\Kp data
suggests that these galaxies have a dominant population that formed
at $z \gtrsim 5$. Detailed morphological and
spectroscopic studies of clusters at high redshift confirm that these
galaxies are a subset of the clusters' elliptical population (e.g. van Dokkum \etal 1999, SED98). However, because of
the non-uniqueness of the spectral synthesis model, it is important
to differentiate between a stellar population that is predominantly
old and one that is entirely old. Although models with recent
formation redshifts (i.e. galaxies with stellar
populations that are entirely formed at $z < 2$)
do not fit the color envelope data, it is not evident from the
previous discussion whether or not models with some recent star formation superposed
on an older population fit the data. 

The observed behavior of the luminosity function suggests significant
evolution in the cluster galaxy population. Because of our relatively
bright magnitude limit, this result pertains almost entirely to
luminous galaxies ($L > L_*$) that primarily become (or have always
been) elliptical galaxies in the cores of these rich
clusters. Therefore, on the one hand we have a seemingly quiescent
(although not necessarily quiescent) elliptical population and on the
other an evolving population. Is it possible that the entire
population is evolving? One solution is suggested by the observation
of red mergers in MS1054$-$03 by van Dokkum \etal (1999). These galaxies
will remain on the red sequence after the merger but 
$M^{*}$ will be affected. 

\begin{figure*}
\figurenum{16}
\plotone{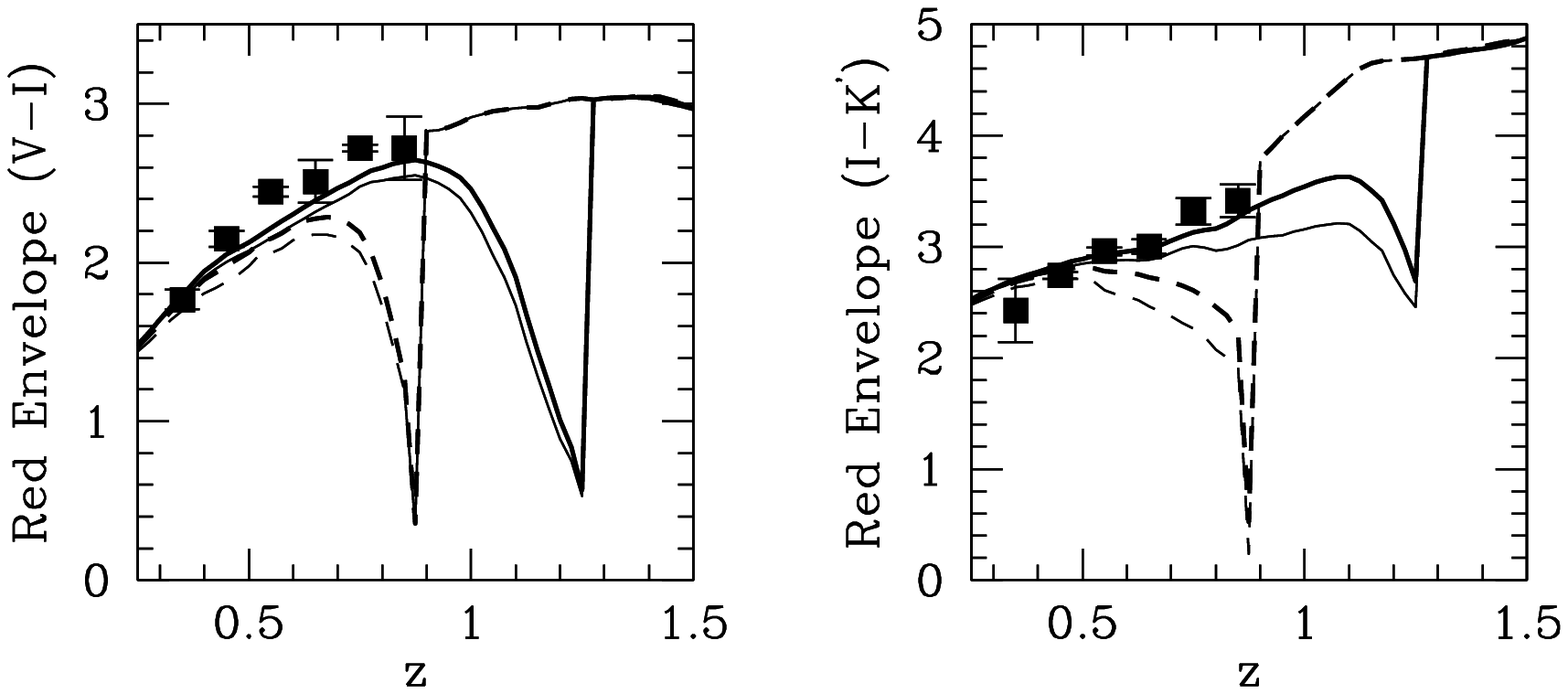}
\protect\figcaption{The observed location of the red envelope in $V-I$
(\textit{left panel}) and $I-$\Kp (\textit{right panel}) compared to
spectral synthesis models of Bruzual \& Charlot (1993) for a
passively evolving elliptical galaxy that includes a secondary burst
of star formation.  The data are binned
in redshift intervals of $\Delta z = 0.1$ and have errors bars which
are $\sigma_{mean}$ of the binned data.  In all models, the elliptical forms at
$z_{form} = 10$ and passively evolves until experiencing a secondary
burst of star formation at $z_{burst} = 1.5$ (\textit{solid lines}) or $z_{burst}
= 1$ (\textit{dashed lines}).  We investigate burst strengths (the
percentage of the galaxy's mass that is converted into stars) of 10\%
(\textit{thick lines}) and 25\% (\textit{thin lines}).}
\label{fig:redburst}
\end{figure*}

Can mergers that do
induce some star formation also account for the observations?  In
Figure~\ref{fig:redburst} we compare the observed evolution of the red envelope
in $V-I$ (\textit{left panel}) and $I-$\Kp (\textit{right panel}), and
spectral synthesis models (Bruzual \& Charlot 1993) of starburst
episodes in elliptical galaxies.  These models consist of a passively
evolving elliptical galaxy that formed at $z_{form}=10$ and
subsequently underwent a burst of star formation at low redshift,
$z_{burst}=1.5$ (\textit{solid lines}) and $z_{burst}=1$
(\textit{dashed lines}).  The duration of the burst is 10$^{7}$ years and has
strengths (amount of galaxy's mass that is converted into stars) of
10\% (\textit{thick lines}) and 25\% (\textit{thin lines}).  At the peak of the
starburst, the bluening of the galaxy's integrated light is dramatic
($\sim$2.5 magnitudes in $V-I$), but also short-lived.  The bulk of
the post-starburst period involves a moderate blueing of the galaxy
($\sim0.2$-1 magnitude in $V-I$).  The strength of the burst does not
significantly affect the amount of blueing, but does affect its
duration.  It is evident that an elliptical galaxy returns fairly
quickly to its standard location on the red E/SO sequence and that the
galaxy appears similar to one that simply had a single burst of star
formation at $z = 10$. We stress that as galaxies evolution they may
or may not be included in subpopulations like the red sequence, which
leads to complicated evolutionary effects (see van Dokkum and Franx
2001 for a more complete discussion of such matters).  It is generally
thought that galaxies primarily move onto the red sequence from a
bluer part of the color-magnitude space (cf. Kodama \& Bower 2001 for
a recent treatment), we simply note here that the flow may go both
ways.

Detailed spectroscopic observations and analysis of one high-redshift cluster
(MS1054-03 at $z=0.83$; van Dokkum \etal 1999) provides observational
confirmation of this hypothesis. They demonstrate that
17\% of the L $\gtrsim$ L$_{*}$ cluster
population is undergoing a major merger. 
Most of these galaxies will probably
evolve into luminous ($\sim$2L$_{*}$) elliptical galaxies.
Assuming that the galaxy population of MS1054-03 is representative of
clusters at that redshift, van Dokkum \etal(1999) estimate that
$\sim$50\% of present-day cluster ellipticals were assembled in
mergers since $z \sim 1$.  Most of the mergers involve red spheroidal
galaxies with no detected [OII] $\lambda$3727 emission, but some do
exhibit enhanced Balmer absorption indicative of a modest recent
starburst.  They measure the scatter in the rest-frame $U-V$ colors of
morphologically selected ellipticals and S0s to be $0.027 \pm 0.013$,
while that of the combined sample of ellipticals and mergers is
significantly higher, $0.054 \pm 0.011$. Evolving the CM relation,
the scatter of the combined sample drops to $\sim 0.035$ at $z=0.5$
and $\sim 0.015$ at $z=0$ which is consistent with the small scatter
observed in the CM relation at low redshift (Bower, Lucey, \& Ellis
1992; Ellis \etal 1997; Stanford \etal 1998; van Dokkum \etal 1998). Using the colors of the mergers they estimated their
luminosity-weighted mean redshift of formation is $z_{form}\gtrsim 1.7$,
which is in remarkable agreement with our $z_{form}$ derived from the
evolution of \Mstar. 

The triggering of star formation in otherwise red E/SO galaxies has
a few consequences: 1) it will weaken or destroy the distinctiveness
of the E/SO sequence in the CM diagram and 2) it will systematically brighten the
population of ellipticals throughout the duration of the burst. Both
of these trends are observed in our data, and in previous
observations. Morphologically, the blue galaxies appear to be
predominantly disky (Couch \etal 1994, 1998; Dressler \etal 1994,
1999; Oemler, Dressler, \& Butcher 1997; Poggianti \etal 1999), but
whether these are spirals evolving to S0's and/or S0's evolving to E's
is not addressed by our data.  A further complication, whether these
evolutionary processes are driven/influenced/terminated by the cluster
environment, is again not addressable with our data because we have no
proper field comparison sample.

Alternatively, one could hypothesize that a significant fraction of
the galaxies on current-day E/S0 cluster sequences came not from
previous early-type galaxies, but from some major morphological evolution
of later types (i.e. spiral-spiral mergers). However, the tightness
of the red sequence (cf. Smail \etal 1998) and the detailed analysis
of galaxy spectra from a sample of nine clusters at $0.37 \leq
z \leq 0.56$ (Jones, Smail, \& Couch 2000) demonstrate that these galaxies
could not have experienced a major episode of star formation in their
relatively recent past. The latter study concludes that the progenitors
of early-type cluster galaxies are themselves early-type cluster
galaxies.

In closing, our results do not drastically alter trends seen previously
from smaller cluster samples. Instead, they provide better
sampled data with which to measure evolutionary trends in cluster 
galaxies for redshifts
from 0.3 to 0.9. The large sample provides confirmation that these 
trends are not dominated by a one or two ``odd" clusters at high redshift.
The empirical results are straightforward (albeit open to a range of
interpretations) and provide tests of theoretical models. 

\section{Summary}

We analyze photometry and spectroscopy of a sample of 63 clusters at
0.3 $\leq z \leq$ 0.9 drawn
from the Las Campanas Distant Cluster Survey. Seventeen of our clusters have
spectroscopically confirmed redshifts and the remaining 46 have
photometrically estimated redshifts. Using deep $V$, $I$, and \Kp
follow-up imaging we measure the luminosity and color evolution of cluster
galaxies in order to empirically constrain models of cluster
galaxy formation and evolution. Our primary results are:

\noindent 1) We develop three photometric redshift estimation techniques based
upon the magnitude of the brightest cluster galaxy measured directly
from our drift scan images, the $I$- band luminosity function, and the
location of the red envelope in both $V-I$ and $I-$\Kp. All three
techniques work well with internally estimated errors of $\sigma_{z_{BCG}}
= 0.05$, $\sigma_{lum} = 0.06$, and $\sigma_{col} = 0.07$. These uncertainties
are derived from the spectroscopic sample and may not represent the 
uncertainties for the entire catalog.

\noindent 2) We find no systematic difference in the photometric properties
(i.e. luminosity function, location of the red envelope, fraction of
blue galaxies) of our spectroscopically confirmed clusters and the
rest of the sample. Nor do we find any
systematic differences between the properties of our clusters and those published in the
literature. Both of these factors lead us to believe that our cluster
finding algorithm and our cluster classification criteria are
successful.

\noindent 3) Using 44 of
our clusters and 19 clusters from the literature (A93; S97; SED98) for
which $I-$band data are available,
we parameterize
the redshift dependence 
of $M^*_I$ in the observed frame: 
$M^*_I = (-21.74\pm0.12) - (0.88\pm0.24)z - 5 \log h$ for 
$0.3 \leq z \leq 0.9$, 

\noindent 4) By combining the $V$ and $I$ photometry of 30 of our clusters and 14
clusters from the literature (A93; S97; SED98) we parameterize
the redshift dependence of the $V-I$ color of the E/S0 red sequence in the
observed frame:
$(V-I) = (-0.24\pm0.28) + (7.42\pm1.03)z - (4.61\pm0.91)z^2$ for $0.3 \leq z \leq 0.9$, 

\noindent 5) By combining 
13 of our clusters with 15 clusters from the literature
(A93; SED98) for which 
$I$ and $K^\prime$ data are available, we parameterize 
the redshift dependence of the $I-$\Kp color of the E/S0 red sequence
in the observed frame:
$(I-$\Kp$) = (0.66\pm0.65) + (9.50\pm3.72)z - (14.72\pm7.01)z^2 + (8.72\pm4.29)z^3$

\noindent 6) Using the peak surface brightness of the cluster
detection fluctuation, $\Sigma$, as a proxy for cluster L$_{x}$, we
find no correlation between $\Sigma$ and (\Mstar$+5 \log h)$ or the
location of the red envelope in $(V-I)$.

\noindent 7) We speculate that all of these observations
can be explained with a model in which luminous
early type galaxies (or the progenitors
of current day early type galaxies) form the bulk of their stellar
populations at high redshifts ($\gtrsim 5$) and in which many of these
galaxies, if not all,
experience a short term episode of star formation at lower redshifts
($1.5 < z < 2)$. 

This sample represents a significant increase in the number of known
high redshift clusters. Although our data are too crude to verify if
our proposed evolutionary scenario is correct, the statistical nature
of this study provides convincing evidence that the observed trends
are universal in massive clusters, not the result of a few
outliers. The next step is to utilize the large sample to select
several representative clusters of different masses and environments
at various epochs to investigate in detail the morphologies and star
formation rates of their member galaxies.

\vskip 1in
\noindent
Acknowledgments: The authors wish to thank Caryl Gronwall for
providing the GISSEL96 code and offering valuable advice on its
implementation. We also thank Adam Stanford for generously providing
us with his data. AEN and AHG acknowledge funding from a NSF
grant (AST-9733111). AEN gratefully acknowledges financial support from
the University of California Graduate Research Mentorship Fellowship
program. AHG acknowledges funding from the National Science
Foundation Graduate Research Fellowship Program and the ARCS
Foundation. DZ acknowledges financial support from a
NSF CAREER grant (AST-9733111)
a David and Lucile Packard Foundation
Fellowship, and an Alfred P. Sloan Fellowship. This work was partially supported
by NASA through grant number GO-07327.01-96A from the Space
Telescope Science Institute, which is operated by the Association
of Universities for Research in Astronomy, Inc., under NASA contract
NAS5-26555.

\vfill\eject
\clearpage

\clearpage
\clearpage

\clearpage

\clearpage

\clearpage

\clearpage

\clearpage

\begin{figure*}
\figurenum{7}
\plotone{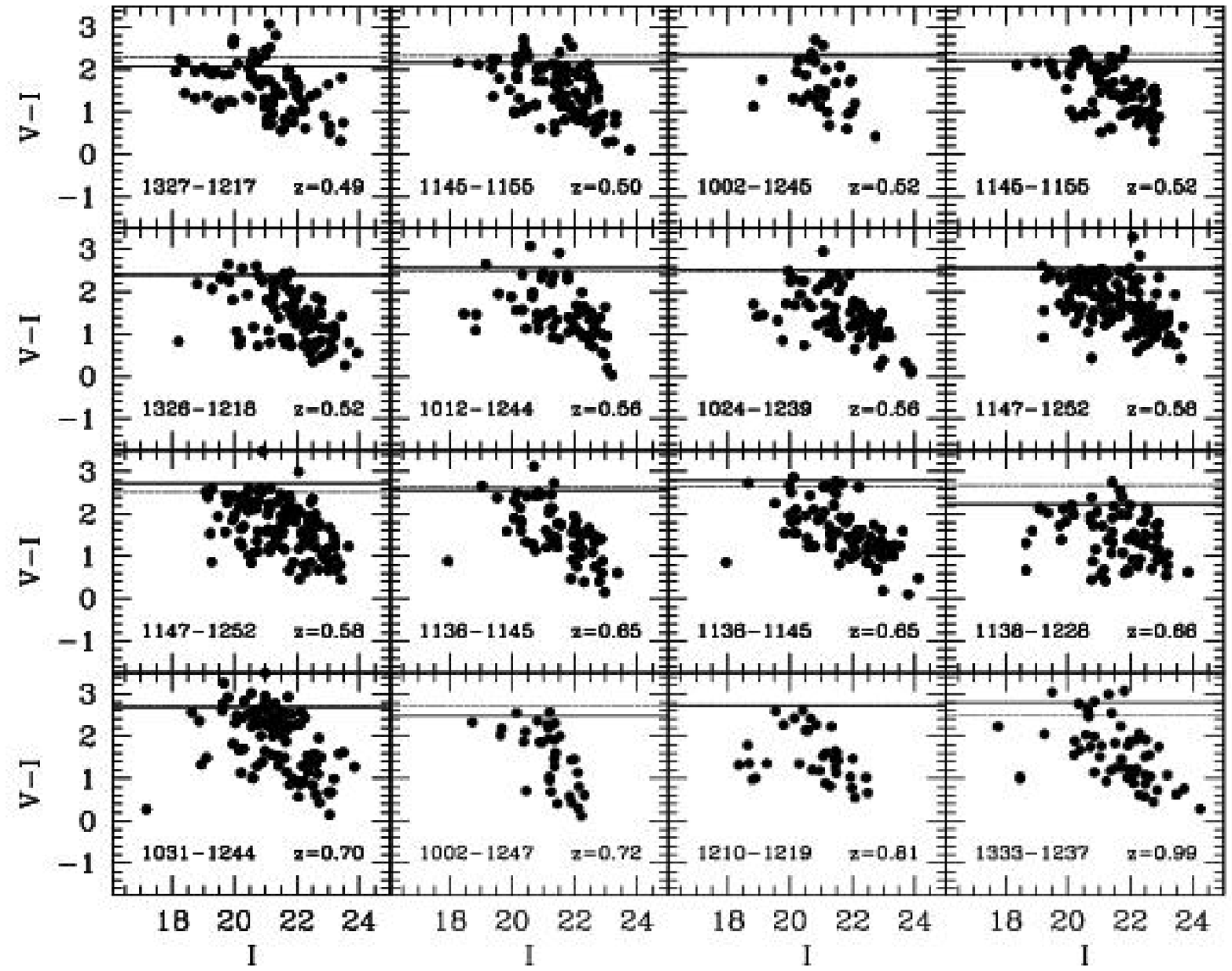}
\protect\figcaption{continued.}
\end{figure*}
\clearpage

\clearpage

\clearpage

\clearpage

\clearpage

\clearpage

\clearpage

\clearpage

\clearpage

\clearpage
\begin{deluxetable}{cccccccccccc}
\tablenum{1}
\tablecaption{Observations}
\tablehead{
\colhead{} & \colhead{RA} & \colhead{DEC} & \colhead{V} & \colhead{Seeing} & \colhead{I} & \colhead{Seeing} & \colhead{K$^{\prime}$} & \colhead{Seeing} \\ 
\colhead{Cluster} & \colhead{(JD2000)} & \colhead{(JD2000)} & \colhead{(min)} & \colhead{($^{\prime\prime}$)} & \colhead{(min)} & \colhead{($^{\prime\prime}$)} & \colhead{(min)} & \colhead{($^{\prime\prime}$)} }

 \startdata 
0915+4738 & 09:15:51.94 & +47:38:19.9 & 50\tablenotemark{2} & 1.8 & 33\tablenotemark{2} & 1.9 & 56\tablenotemark{2} & 2.0\\
0936+4620 & 09:36:06.28 & +46:20:43.2 & 80\tablenotemark{2} & 2.5 & 80\tablenotemark{2} & 1.9 & 64\tablenotemark{2} & 2.0\\
0944+4732 & 09:44:21.83 & +47:32:42.6 & ... & ... & ... & ... & 40\tablenotemark{4} & 2.0\\
1002$-$1245 & 10:02:01.47 & $-$12:45:35.1 & 120\tablenotemark{1} & 1.4 & 60\tablenotemark{1} & 1.4 & ... & ...\\
1002$-$1247 & 10:02:27.14 & $-$12:47:13.1 & 120\tablenotemark{1} & 1.4 & 60\tablenotemark{1} & 1.4 & ... & ...\\
1005$-$1147 & 10:05:43.60 & $-$11:47:43.1 & ... & ... & 60\tablenotemark{1} & 1.2 & ... & ...\\
1005$-$1209 & 10:05:49.72 & $-$12:09:36.5 & ... & ... & 45\tablenotemark{1} & 1.4 & ... & ...\\
1006$-$1222 & 10:06:29.25 & $-$12:22:13.7 & ... & ... & 45\tablenotemark{1} & 1.4 & ... & ...\\
1006$-$1258 & 10:06:18.79 & $-$12:58:12.5 & ... & ... & 68\tablenotemark{1} & 1.1 & ... & ...\\
1007$-$1208 & 10:07:42.60 & $-$12:08:36.0 & ... & ... & 40\tablenotemark{1} & 1.3 & ... & ...\\
1012$-$1243 & 10:12:14.58 & $-$12:43:10.9 & ... & ... & 45\tablenotemark{3} & 0.8 & 32\tablenotemark{3} & 0.8\\
1012$-$1245 & 10:12:44.35 & $-$12:45:37.9 & 30\tablenotemark{3} & 0.9 & 30\tablenotemark{3} & 0.7 & ... & ...\\
1014$-$1143 & 10:14:56.31 & $-$11:43:08.1 & ... & ... & 45\tablenotemark{1} & 1.2 & ... & ...\\
1015$-$1132 & 10:15:19.47 & $-$11:32:55.5 & ... & ... & 60\tablenotemark{1} & 1.3 & ... & ...\\
1017$-$1128 & 10:17:45.31 & $-$11:28:07.8 & ... & ... & 60\tablenotemark{3} & 1.1 & ... & ...\\
1018$-$1211 & 10:18:46.45 & $-$12:11:52.8 & 40\tablenotemark{3} & 1.2 & 40\tablenotemark{3} & 0.6 & ... & ...\\
1023$-$1303 & 10:23:10.11 & $-$13:03:52.0 & ... & ... & 30\tablenotemark{3} & 0.6 & ... & ...\\
1024$-$1239 & 10:24:44.94 & $-$12:39:55.6 & 40\tablenotemark{3} & 1.1 & 40\tablenotemark{3} & 0.9 & ... & ...\\
1025$-$1236 & 10:25:08.83 & $-$12:36:20.0 & 90\tablenotemark{1} & 1.5 & 115\tablenotemark{1} & 2.7 & ... & ...\\
1027$-$1159 & 10:27:26.31 & $-$11:59:33.6 & 90\tablenotemark{1} & 1.5 & 105\tablenotemark{1} & 1.8 & ... & ...\\
1031$-$1244 & 10:31:50.26 & $-$12:44:27.2 & 40\tablenotemark{3} & 1.0 & 60\tablenotemark{3} & 0.8 & ... & ...\\
1032$-$1229 & 10:32:04.91 & $-$12:29:43.8 & 120\tablenotemark{1} & 1.8 & 115\tablenotemark{1} & 1.8 & ... & ...\\
... & ... & ... & ... & ... & 15\tablenotemark{3} & 1.0 & ... & ...\\
1041+4626 & 10:41:03.79 & +46:26:36.3 & 40\tablenotemark{2} & 2.5 & 60\tablenotemark{2} & 2.2 & 118\tablenotemark{2} & 2.0\\
1059+4737 & 10:59:38.03 & +47:37:38.6 & 50\tablenotemark{2} & 3.3 & 50\tablenotemark{2} & 2.0 & 55\tablenotemark{2} & 2.0\\
1100+4620 & 11:00:57.36 & +46:20:38.3 & ... & ... & ... & ... & 88\tablenotemark{2} & 2.0\\
1136$-$1136 & 11:36:31.95 & $-$11:36:07.7 & 30\tablenotemark{3} & 1.3 & 30\tablenotemark{3} & 1.1 & ... & ...\\
... & ... & ... & ... & ... & 20\tablenotemark{1} & 1.1 & ... & ...\\
1136$-$1145 & 11:36:47.07 & $-$11:45:33.3 & 13\tablenotemark{3} & 1.1 & 20\tablenotemark{3} & 0.9 & 44\tablenotemark{3} & 0.8\\
... & ... & ... & 40\tablenotemark{3} & 1.0 & 40\tablenotemark{3} & 0.8 & ... & ...\\
1136$-$1252 & 11:36:33.54 & $-$12:52:03.3 & 20\tablenotemark{3} & 0.8 & 20\tablenotemark{3} & 0.6 & 20\tablenotemark{4} & 2.0\\
... & ... & ... & ... & ... & 20\tablenotemark{3} & 1.4 & ... & ...\\
1138$-$1142 & 11:38:06.59 & $-$11:42:10.3 & 20\tablenotemark{3} & 1.1 & 30\tablenotemark{3} & 0.7 & 32\tablenotemark{3} & 0.8\\
... & ... & ... & 60\tablenotemark{3} & 1.3 & 60\tablenotemark{3} & 1.2 & ... & ...\\
1138$-$1225 & 11:38:14.24 & $-$12:25:53.9 & 90\tablenotemark{1} & 1.2 & 120\tablenotemark{1} & 2.0 & ... & ...\\
1138$-$1228 & 11:38:44.70 & $-$12:28:25.4 & 30\tablenotemark{3} & 0.7 & 20\tablenotemark{3} & 0.8 & ... & ...\\
1139$-$1154 & 11:39:30.47 & $-$11:54:25.0 & 30\tablenotemark{3} & 0.9 & 30\tablenotemark{3} & 0.8 & 10\tablenotemark{4} & 2.0\\
1139$-$1217 & 11:39:56.84 & $-$12:17:19.8 & 20\tablenotemark{3} & 0.7 & 20\tablenotemark{3} & 0.8 & 32\tablenotemark{3} & 0.8\\
1145$-$1155 & 11:45:22.36 & $-$11:55:52.1 & 20\tablenotemark{3} & 0.7 & 20\tablenotemark{3} & 1.0 & 50\tablenotemark{4} & 2.0\\
... & ... & ... & 40\tablenotemark{3} & 1.2 & 30\tablenotemark{3} & 0.8 & ... & ...\\
1147$-$1252 & 11:47:16.98 & $-$12:52:04.7 & 20\tablenotemark{3} & 1.0 & 20\tablenotemark{3} & 0.9 & 32\tablenotemark{3} & 0.8\\
... & ... & ... & 40\tablenotemark{3} & 1.2 & 30\tablenotemark{3} & 0.7 & ... & ...\\
1149$-$1159 & 11:49:22.14 & $-$11:59:19.3 & 3\tablenotemark{3} & 0.9 & 10\tablenotemark{3} & 0.9 & ... & ...\\
1149$-$1246 & 11:49:06.18 & $-$12:46:06.3 & 20\tablenotemark{3} & 0.9 & 20\tablenotemark{3} & 0.8 & 32\tablenotemark{3} & 0.8\\
... & ... & ... & 40\tablenotemark{3} & 1.0 & 40\tablenotemark{3} & 0.7 & ... & ...\\
1208$-$1151 & 12:08:26.67 & $-$11:51:25.6 & 90\tablenotemark{1} & 1.4 & 68\tablenotemark{1} & 1.2 & ... & ...\\
1210$-$1219 & 12:10:12.73 & $-$12:19:06.9 & 60\tablenotemark{1} & 1.4 & 60\tablenotemark{1} & 1.4 & ... & ...\\
\enddata

\end{deluxetable}

\clearpage
\begin{deluxetable}{cccccccccccc}
\tablenum{1}
\tablecaption{Observations: Continued}
\tablehead{
\colhead{} & \colhead{RA} & \colhead{DEC} & \colhead{V} & \colhead{Seeing} & \colhead{I} & \colhead{Seeing} & \colhead{K$^{\prime}$} & \colhead{Seeing} \\ 
\colhead{Cluster} & \colhead{(JD2000)} & \colhead{(JD2000)} & \colhead{(min)} & \colhead{($^{\prime\prime}$)} & \colhead{(min)} & \colhead{($^{\prime\prime}$)} & \colhead{(min)} & \colhead{($^{\prime\prime}$)} }

 \startdata 
1211$-$1220 & 12:11:04.16 & $-$12:20:47.7 & ... & ... & 45\tablenotemark{3} & 0.8 & ... & ...\\
1215$-$1252 & 12:15:41.08 & $-$12:52:59.7 & ... & ... & 68\tablenotemark{1} & 1.1 & ... & ...\\
1216$-$1201 & 12:16:45.10 & $-$12:01:17.3 & ... & ... & 45\tablenotemark{1} & 1.3 & ... & ...\\
1219$-$1154 & 12:19:34.88 & $-$11:54:22.9 & ... & ... & 60\tablenotemark{1} & 1.3 & ... & ...\\
1219$-$1201 & 12:19:44.49 & $-$12:01:35.5 & ... & ... & 60\tablenotemark{1} & 1.3 & ... & ...\\
1221$-$1206 & 12:21:46.20 & $-$12:06:12.7 & ... & ... & 45\tablenotemark{1} & 1.2 & ... & ...\\
1230+4621 & 12:30:16.26 & +46:21:17.1 & 67\tablenotemark{2} & 3.8 & 33\tablenotemark{2} & 2.2 & 72\tablenotemark{2} & 2.0\\
1326$-$1218 & 13:26:12.66 & $-$12:18:22.5 & 40\tablenotemark{3} & 0.9 & 50\tablenotemark{3} & 0.9 & 36\tablenotemark{3} & 0.8\\
1327$-$1217 & 13:27:57.33 & $-$12:17:16.7 & 20\tablenotemark{3} & 1.1 & 20\tablenotemark{3} & 1.0 & 42\tablenotemark{3} & 0.8\\
... & ... & ... & 40\tablenotemark{3} & 1.3 & 50\tablenotemark{3} & 0.8 & ... & ...\\
1329$-$1256 & 13:29:11.37 & $-$12:56:22.0 & 90\tablenotemark{1} & 1.6 & 95\tablenotemark{1} & 1.4 & 20\tablenotemark{4} & 2.0\\
... & ... & ... & 45\tablenotemark{3} & 1.1 & 30\tablenotemark{3} & 1.0 & ... & ...\\
1333$-$1237 & 13:33:01.92 & $-$12:37:17.0 & 20\tablenotemark{3} & 0.9 & 30\tablenotemark{3} & 1.2 & ... & ...\\
1404$-$1216 & 14:04:47.20 & $-$12:16:21.4 & 90\tablenotemark{1} & 1.5 & 60\tablenotemark{1} & 1.3 & ... & ...\\
1405$-$1147 & 14:05:11.38 & $-$11:47:08.6 & ... & ... & 60\tablenotemark{1} & 1.4 & ... & ...\\
1406$-$1232 & 14:06:36.54 & $-$12:32:39.7 & ... & ... & 45\tablenotemark{1} & 1.4 & ... & ...\\
1408$-$1209 & 14:08:17.86 & $-$12:09:27.0 & ... & ... & 68\tablenotemark{1} & 1.0 & ... & ...\\
1408$-$1216 & 14:08:45.95 & $-$12:16:08.8 & ... & ... & 68\tablenotemark{1} & 1.0 & ... & ...\\
1408$-$1218 & 14:08:50.62 & $-$12:18:14.8 & ... & ... & 68\tablenotemark{1} & 1.0 & ... & ...\\
1412$-$1150 & 14:12:32.56 & $-$11:50:16.2 & ... & ... & 8\tablenotemark{1} & 1.2 & ... & ...\\
... & ... & ... & ... & ... & 30\tablenotemark{3} & 0.8 & ... & ...\\
1412$-$1222 & 14:12:25.92 & $-$12:22:53.3 & ... & ... & 8\tablenotemark{1} & 1.3 & ... & ...\\
... & ... & ... & ... & ... & 45\tablenotemark{3} & 0.8 & ... & ...\\
1412$-$1222.1 & 14:12:31.26 & $-$12:22:15.8 & ... & ... & 45\tablenotemark{3} & 0.8 & ... & ...\\
1413$-$1244 & 14:13:08.59 & $-$12:44:11.6 & ... & ... & 40\tablenotemark{1} & 1.2 & ... & ...\\
1416$-$1143 & 14:16:45.13 & $-$11:43:40.0 & ... & ... & 40\tablenotemark{1} & 1.3 & ... & ...\\
1422+4622 & 14:22:24.18 & +46:22:39.7 & 80\tablenotemark{2} & 1.7 & 83\tablenotemark{2} & 2.4 & 35\tablenotemark{4} & 2.0\\
1519+4622 & 15:19:54.27 & +46:22:20.4 & ... & ... & 15\tablenotemark{2} & 2.3 & 45\tablenotemark{4} & 2.0
\enddata
\vfill\eject
\tablenotetext{(1)}{Las Campanas 1m}
\tablenotetext{(2)}{Palomar 1.5m}
\tablenotetext{(3)}{Las Campanas 2.5m}
\tablenotetext{(4)}{Lick 3m}
\end{deluxetable}

\clearpage
\begin{deluxetable}{cccccccccccccc}
\tablenum{2}
\tablecaption{Redshift Data}
\tablehead{
\colhead{Cluster} & \colhead{z$_{spec}$} & \colhead{z$_{lum}$} & \colhead{$\sigma_{z_{lum}}$} & \colhead{z$_{col}$} & \colhead{$\sigma_{z_{col}}$} & \colhead{z$_{bcg}$} & \colhead{$\sigma_{z_{bcg}}$} & \colhead{m$^{*}_{I}$} & \colhead{$\sigma_{m^{*}_{I}}$} & \colhead{V$-$I} & \colhead{$\sigma_{V-I}$} & \colhead{I$-$K} & \colhead{$\sigma_{I-K}$} } 

 \startdata 
0915+4738 & 0.40 & 0.43 & 0.06 & 0.50 & 0.07 & ... & ... & 18.77 & 0.08 & 2.12 & 0.05 & 2.88 & 0.03\\
0936+4620 & 0.54 & 0.49 & 0.06 & ... & ... & ... & ... & 18.97 & 0.08 & ... & ... & 2.96 & 0.03\\
0944+4732 & 0.58 & ... & ... & ... & ... & ... & ... & ... & ... & ... & ... & ... & ...\\
1002$-$1245 & 0.52 & 0.64 & 0.06 & 0.50 & 0.07 & 0.43 & 0.05 & 19.65 & 0.11 & 2.32 & 0.03 & ... & ...\\
1002$-$1247 & ... & 0.72 & 0.06 & 0.55 & 0.07 & 0.57 & 0.05 & 19.82 & 0.14 & 2.48 & 0.03 & ... & ...\\
1005$-$1147 & ... & 0.89 & 0.06 & ... & ... & 0.53 & 0.05 & 20.24 & 0.21 & ... & ... & ... & ...\\
1005$-$1209 & ... & 0.45 & 0.06 & ... & ... & ... & ... & 18.82 & 0.08 & ... & ... & ... & ...\\
1006$-$1222 & ... & 1.01 & 0.06 & ... & ... & 0.72 & 0.05 & 20.54 & 0.21 & ... & ... & ... & ...\\
1006$-$1258 & ... & 0.73 & 0.06 & ... & ... & 0.47 & 0.05 & 19.88 & 0.14 & ... & ... & ... & ...\\
1007$-$1208 & ... & 0.49 & 0.06 & ... & ... & 0.41 & 0.05 & 18.99 & 0.08 & ... & ... & ... & ...\\
1012$-$1243 & ... & 0.49 & 0.06 & ... & ... & 0.44 & 0.05 & 18.99 & 0.08 & ... & ... & ... & ...\\
1012$-$1245 & ... & 0.56 & 0.06 & 0.59 & 0.07 & ... & ... & 19.28 & 0.07 & 2.56 & 0.07 & ... & ...\\
1014$-$1143 & ... & 0.30 & 0.06 & ... & ... & ... & ... & 17.95 & 0.12 & ... & ... & ... & ...\\
1015$-$1132 & ... & 0.60 & 0.06 & ... & ... & 0.45 & 0.05 & 19.43 & 0.07 & ... & ... & ... & ...\\
1017$-$1128 & ... & 0.50 & 0.06 & ... & ... & 0.66 & 0.05 & 19.04 & 0.08 & ... & ... & ... & ...\\
1018$-$1211 & ... & 0.48 & 0.06 & 0.41 & 0.07 & 0.47 & 0.05 & 18.94 & 0.08 & 2.00 & 0.07 & ... & ...\\
1023$-$1303 & ... & 0.51 & 0.06 & ... & ... & 0.77 & 0.05 & 19.10 & 0.07 & ... & ... & ... & ...\\
1024$-$1239 & ... & 0.56 & 0.06 & 0.57 & 0.07 & 0.76 & 0.05 & 19.28 & 0.07 & 2.52 & 0.03 & ... & ...\\
1025$-$1236 & ... & 0.38 & 0.06 & 0.35 & 0.07 & 0.34 & 0.05 & 18.48 & 0.12 & 1.68 & 0.07 & ... & ...\\
1027$-$1159 & ... & 0.30 & 0.06 & 0.37 & 0.07 & 0.37 & 0.05 & 17.95 & 0.12 & 1.80 & 0.07 & ... & ...\\
1031$-$1244 & ... & 0.70 & 0.06 & 0.65 & 0.07 & 0.70 & 0.05 & 19.76 & 0.11 & 2.68 & 0.07 & ... & ...\\
1032$-$1229 & ... & 0.44 & 0.06 & 0.41 & 0.07 & 0.45 & 0.05 & 18.76 & 0.08 & 1.96 & 0.07 & ... & ...\\
          & ... & 0.50 & 0.06 & ... & ... & ... & ... & 19.04 & 0.08 & ... & ... & ... & ...\\
1041+4626 & 0.62 & 0.61 & 0.06 & ... & ... & ... & ... & 19.45 & 0.11 & ... & ... & 3.04 & 0.07\\
1059+4737 & 0.36 & 0.33 & 0.06 & 0.28 & 0.07 & ... & ... & 18.18 & 0.12 & 1.52 & 0.07 & 3.00 & 0.07\\
1100+4620 & 0.46 & ... & ... & ... & ... & ... & ... & ... & ... & ... & ... & ... & ...\\
1136$-$1136 & ... & 0.34 & 0.06 & 0.42 & 0.07 & ... & ... & 18.20 & 0.12 & 2.04 & 0.05 & ... & ...\\
          & ... & 0.34 & 0.06 & ... & ... & ... & ... & 18.24 & 0.12 & ... & ... & ... & ...\\
1136$-$1145 & 0.65 & 0.59 & 0.06 & 0.59 & 0.07 & 0.67 & 0.05 & 19.31 & 0.07 & 2.56 & 0.07 & 2.92 & 0.03\\
          & ...  & 0.67 & 0.06 & 0.77 & 0.07 & ... & ... & 19.68 & 0.11 & 2.80 & 0.21 & 2.96 & 0.03\\
1136$-$1252 & ... & 0.39 & 0.06 & 0.42 & 0.07 & 0.37 & 0.05 & 18.49 & 0.12 & 2.04 & 0.05 & ... & ...\\
          & ... & 0.31 & 0.06 & ... & ... & ... & ... & 18.00 & 0.12 & ... & ... & ... & ...\\
1138$-$1142 & 0.45 & 0.45 & 0.06 & 0.51 & 0.07 & 0.52 & 0.05 & 18.81 & 0.08 & 2.36 & 0.03 & 2.72 & 0.04\\
          & ... & 0.46 & 0.06 & 0.54 & 0.07 & ... & ... & 18.86 & 0.08 & 2.44 & 0.03 & ... & ...\\
1138$-$1225 & ... & 0.36 & 0.06 & 0.30 & 0.07 & 0.36 & 0.05 & 18.34 & 0.12 & 1.44 & 0.07 & ... & ...\\
1138$-$1228 & ... & 0.66 & 0.06 & 0.48 & 0.07 & ... & ... & 19.63 & 0.11 & 2.24 & 0.05 & ... & ...\\
1139$-$1154 & ... & 0.44 & 0.06 & 0.41 & 0.07 & ... & ... & 18.76 & 0.08 & 1.96 & 0.07 & ... & ...\\
1139$-$1217 & 0.48 & 0.44 & 0.06 & 0.49 & 0.07 & ... & ... & 18.70 & 0.12 & 2.28 & 0.05 & 2.72 & 0.04\\
1145$-$1155 & ... & 0.52 & 0.06 & 0.47 & 0.07 & 0.44 & 0.05 & 19.13 & 0.07 & 2.20 & 0.05 & ... & ...\\
          & ... & 0.50 & 0.06 & 0.45 & 0.07 & ... & ... & 19.04 & 0.08 & 2.16 & 0.05 & ... & ...\\
1147$-$1252 & 0.58 & 0.53 & 0.06 & 0.59 & 0.07 & 0.62 & 0.05 & 19.05 & 0.08 & 2.56 & 0.07 & 2.92 & 0.03\\
          & ... & 0.49 & 0.06 & 0.69 & 0.07 & ... & ... & 18.93 & 0.08 & 2.72 & 0.03 & 3.00 & 0.07\\
1149$-$1159 & ... & 0.32 & 0.06 & 0.31 & 0.07 & 0.23 & 0.05 & 18.09 & 0.12 & 1.48 & 0.07 & ... & ...\\
1149$-$1246 & 0.48 & 0.45 & 0.06 & 0.47 & 0.07 & 0.48 & 0.05 & 18.85 & 0.08 & 2.20 & 0.05 & 2.72 & 0.04\\
          & ... & 0.52 & 0.06 & 0.51 & 0.07 & ... & ... & 19.15 & 0.07 & 2.36 & 0.03 & 2.76 & 0.04\\
1208$-$1151 & ... & 0.61 & 0.06 & ... & ... & ... & ... & 19.47 & 0.11 & ... & ... & ... & ...\\
\enddata
\end{deluxetable}

\clearpage

\begin{deluxetable}{cccccccccccccc}
\tablenum{2}
\tablecaption{Redshift Data: Continued}
\tablehead{
\colhead{Cluster} & \colhead{z$_{spec}$} & \colhead{z$_{lum}$} & \colhead{$\sigma_{z_{lum}}$} & \colhead{z$_{col}$} & \colhead{$\sigma_{z_{col}}$} & \colhead{z$_{bcg}$} & \colhead{$\sigma_{z_{bcg}}$} & \colhead{m$^{*}_{I}$} & \colhead{$\sigma_{m^{*}_{I}}$} & \colhead{V$-$I} & \colhead{$\sigma_{V-I}$} & \colhead{I$-$K} & \colhead{$\sigma_{I-K}$} } 

 \startdata 
1210$-$1219 & ... & 0.81 & 0.06 & 0.68 & 0.07 & 0.74 & 0.05 & 20.06 & 0.14 & 2.72 & 0.03 & ... & ...\\
1211$-$1220 & ... & 0.89 & 0.06 & ... & ... & 0.52 & 0.05 & 20.24 & 0.21 & ... & ... & ... & ...\\
1215$-$1252 & ... & 0.39 & 0.06 & ... & ... & ... & ... & 18.49 & 0.12 & ... & ... & ... & ...\\
1216$-$1201 & 0.80 & 0.92 & 0.06 & ... & ... & 0.80 & 0.05 & 20.33 & 0.21 & ... & ... & ... & ...\\
1219$-$1154 & ... & 0.83 & 0.06 & ... & ... & 0.68 & 0.05 & 20.13 & 0.14 & ... & ... & ... & ...\\
1219$-$1201 & ... & 0.56 & 0.06 & ... & ... & 0.55 & 0.05 & 19.28 & 0.07 & ... & ... & ... & ...\\
1221$-$1206 & ... & 0.34 & 0.06 & ... & ... & ... & ... & 18.20 & 0.12 & ... & ... & ... & ...\\
1230+4621 & 0.51 & ... & ... & ... & ... & ... & ... & ... & ... & ... & ... & 3.08 & 0.21\\
1326$-$1218 & ... & 0.52 & 0.06 & 0.53 & 0.07 & ... & ... & 19.12 & 0.07 & 2.40 & 0.03 & 2.72 & 0.04\\
1327$-$1217 & ... & 0.41 & 0.06 & 0.43 & 0.07 & ... & ... & 18.60 & 0.12 & 2.08 & 0.05 & 2.80 & 0.04\\
          & ... & 0.49 & 0.06 & 0.43 & 0.07 & ... & ... & 18.99 & 0.08 & 2.08 & 0.05 & 2.56 & 0.29\\
1329$-$1256 & ... & 0.31 & 0.06 & 0.37 & 0.07 & 0.37 & 0.05 & 18.00 & 0.12 & 1.80 & 0.07 & 2.16 & 0.29\\
          & ... & 0.34 & 0.06 & 0.36 & 0.07 & ... & ... & 18.20 & 0.12 & 1.76 & 0.07 & 2.12 & 0.29\\
1333$-$1237 & ... & 0.99 & 0.06 & 0.77 & 0.07 & ... & ... & 20.50 & 0.21 & 2.80 & 0.21 & ... & ...\\
1404$-$1216 & 0.39 & 0.34 & 0.06 & 0.35 & 0.07 & 0.47 & 0.05 & 18.28 & 0.12 & 1.72 & 0.07 & ... & ...\\
1405$-$1147 & ... & 0.96 & 0.06 & ... & ... & 0.64 & 0.05 & 20.42 & 0.21 & ... & ... & ... & ...\\
1406$-$1232 & ... & 0.96 & 0.06 & ... & ... & ... & ... & 20.43 & 0.21 & ... & ... & ... & ...\\
1408$-$1209 & ... & 0.34 & 0.06 & ... & ... & 0.48 & 0.05 & 18.19 & 0.12 & ... & ... & ... & ...\\
1408$-$1216 & ... & 0.60 & 0.06 & ... & ... & ... & ... & 19.43 & 0.07 & ... & ... & ... & ...\\
1408$-$1218 & ... & 0.60 & 0.06 & ... & ... & 0.55 & 0.05 & 19.43 & 0.07 & ... & ... & ... & ...\\
1412$-$1150 & ... & 0.66 & 0.06 & ... & ... & 0.71 & 0.05 & 19.63 & 0.11 & ... & ... & ... & ...\\
          & ... & 0.75 & 0.06 & ... & ... & ... & ... & 19.90 & 0.14 & ... & ... & ... & ...\\
1412$-$1222 & ... & 0.32 & 0.06 & ... & ... & ... & ... & 18.09 & 0.12 & ... & ... & ... & ...\\
          & ... & 0.31 & 0.06 & ... & ... & ... & ... & 18.00 & 0.12 & ... & ... & ... & ...\\
1412$-$1222.1 & ... & 0.58 & 0.06 & ... & ... & ... & ... & 19.32 & 0.07 & ... & ... & ... & ...\\
1413$-$1244 & ... & 0.81 & 0.06 & ... & ... & 0.63 & 0.05 & 20.06 & 0.14 & ... & ... & ... & ...\\
1416$-$1143 & ... & 0.54 & 0.06 & ... & ... & ... & ... & 19.20 & 0.07 & ... & ... & ... & ...\\
1422+4622 & 0.69 & ... & ... & 0.32 & ... & ... & ... & ... & ... & 1.56 & ... & ... & ...\\
1519+4622 & 0.82 & ... & ... & ... & ... & ... & ... & ... & ... & ... & ... & ... & ...
\enddata
\end{deluxetable}

\clearpage
\def\Kp{K$^\prime$\ }
\begin{deluxetable}{lll}
\tablewidth{0pc}
\tablenum{3}
\tablecaption{Evolution Parametrizations}
\tablehead{
\colhead{} & \colhead{Functional} & \colhead{} \\
\colhead{Type} & \colhead{Form} & \colhead{Coefficients} }

\startdata 

%Luminosity & (M$^{*}_{I}+5 \log h)=({\rm M}^{*}_{I_{0}}+5 \log h)+{\rm C}z$ & $({\rm M}^{*}_{I_{0}}+5 \log h)=-21.70\pm0.04$ \\
Luminosity & (M$^{*}_{I}+5 \log h)={\rm C_{1}}+{\rm C_{2}}z$ & C$_{1}=-21.70\pm0.04$ \\
& & C$_{2}=-0.94\pm0.09$ \\
\\
Color ($V-I$) & ($V-I)~=~$C$_{1}+$ C$_{2}z~+~$C$_{3}z^{2}$ & C$_{1}=-0.02\pm0.11$ \\
& & C$_{2}=6.73\pm0.41$ \\
& & C$_{3}=-4.09\pm0.35$ \\
\\
Color ($I-K^{\prime})$ & ($I-K^{\prime})~=~$C$_{1}~+~$C$_{2}z~+~$C$_{3}z^{2}~+~$C$_{4}z^{3}$ & C$_{1}=-0.50\pm0.76$ \\
& & C$_{2}=15.1\pm4.1$ \\
& & C$_{3}=-23.2\pm7.2$ \\
& & C$_{4}=12.6\pm4.2$

\enddata
\end{deluxetable}

\end{document}